\documentclass[aps,prd,10pt,notitlepage,nofootinbib]{revtex4-1}

\usepackage[utf8]{inputenc}
\usepackage{amsmath,amssymb,amsfonts}
\usepackage{newtxtext,newtxmath}

\usepackage{bbold}
\usepackage{bm}
\usepackage{graphicx}
\usepackage[usenames,dvipsnames]{xcolor}
\usepackage{color}
\usepackage[colorlinks=true,linkcolor=blue,urlcolor=blue,citecolor=blue]{hyperref}
\usepackage{slashed}
\usepackage[english]{babel}
\usepackage{dcolumn}
\usepackage{pifont}
\usepackage{dsfont,mathrsfs}
\usepackage{cancel}
\usepackage{bigints}
\usepackage{accents}
\usepackage{soul}
\usepackage{multirow}
\usepackage{bbold}
\usepackage{makecell}
\usepackage{simpler-wick}
\usepackage{setspace}
\usepackage{scalerel}
\usepackage{orcidlink} %Added by Snigdha
\usepackage[normalem]{ulem}
\usepackage{hyperref}

\newcommand{\nn}{\nonumber}

\newcommand{\FB}[1]{\left(#1\right)}
\newcommand{\fb}[1]{(#1)}
\newcommand{\SB}[1]{\left\{#1\right\}}
\newcommand{\TB}[1]{\left[#1\right]}

\newcommand{\munu}{{\mu\nu}}

\newcommand{\RE}{\text{Re}}

\newcommand{\ParOne}[2]{\dfrac{\partial {#1}}{\partial {#2}}}

\newcommand{\fsl}{\slashed}

\newcommand{\unit}{\mathds{1}}  
\newcommand{\sgn}[1]{{\rm sgn} \fb{#1}}

\newcommand{\gm}{\gamma}

\newcommand{\diag}[1]{\texttt{diag}\FB{#1}}
\newcommand{\Tr}[1]{{\rm Tr}\TB{#1}}

\renewcommand{\ap}{a^\prime}
\newcommand{\bp}{b^\prime}
\newcommand{\cp}{c^\prime}

\newcommand{\mF}{\mathcal{F}}
\newcommand{\mA}{\mathcal{A}}
\newcommand{\mO}{\mathcal{O}}
\newcommand{\mP}{\mathcal{P}}
\newcommand{\mN}{\mathcal{N}}
\newcommand{\mS}{\mathcal{S}}

\newcommand{\mhA}{\mathds{A}}
\newcommand{\mhBp}{\mathds{B}^+}
\newcommand{\mhBm}{\mathds{B}^-}

\newcommand{\mtA}{\mathtt{A}}
\newcommand{\mtB}{\mathtt{B}}
\newcommand{\mtC}{\mathtt{C}}
\newcommand{\mtD}{\mathtt{D}}

\newcommand{\fourint}[1]{\int \frac{d^4 {#1}}{(2\pi)^4}}
\newcommand{\threeint}[1]{\int \frac{d^3 {#1}}{(2\pi)^3}}

\newcommand{\Soo}{\Sigma^{11}}
\newcommand{\SooV}{\Sigma^{11}_{\rm vac}}
\newcommand{\Soon}{\Sigma^{11}_{I}}
\newcommand{\Soonn}{\Sigma^{11}_{II}}

\newcommand{\wqmo}{\omega_{q,1}}

\newcommand{\wq}{\omega_q}
\newcommand{\wk}{\omega_k}

\newcommand{\hmN}{\hat{\mN}}
\newcommand{\hmA}{\hat{\mA}}
\newcommand{\hmF}{\hat{\mF}}

\newcommand{\mdPp}{\mathds{P}_+}
\newcommand{\mdPm}{\mathds{P}_-}

\newcommand{\SgB}{\Sigma_B}

\newcommand{\pkcap}{\vec{p}\cdot \hat{k}}
\newcommand{\PKcap}{P\cdot \hat{K}}
\newcommand{\Kcapn}{\hat{K} \cdot n}
\newcommand{\Kcapu}{\hat{K} \cdot u}
\newcommand{\AngInt}{\int \dfrac{d \Omega}{4 \pi }}
\newcommand{\Mth}{M_{\rm th}}

\newcommand{\ZtoInfint}{\int_0^\infty}

\begin{document}
%
%\begin{flushright}
%RBRC-1202
%\end{flushright}

%\global\let\newpage\relax
\title{Collective modes of a massive fermion in a magnetized medium with finite anomalous magnetic moment}

\author{Nilanjan Chaudhuri\orcidlink{0000-0002-7776-3503}$^{a,d}$}
\email{sovon.nilanjan@gmail.com}
\email{n.chaudhri@vecc.gov.in}

\author{Snigdha Ghosh\orcidlink{0000-0002-2496-2007}$^{b}$}
\email{snigdha.ghosh@bangla.gov.in}
%\email{snigdha@ggdckharagpur2.ac.in}
\email{snigdha.physics@gmail.com}
\thanks{Corresponding Author}

\author{Pradip Roy$^{c,d}$}
\email{pradipk.roy@saha.ac.in}

\author{Sourav Sarkar\orcidlink{0000-0002-2952-3767}$^{a,d}$}
\email{sourav@vecc.gov.in}

\affiliation{$^a$Variable Energy Cyclotron Centre, 1/AF Bidhannagar, Kolkata - 700064, India}
\affiliation{$^b$Government General Degree College Kharagpur-II, Paschim Medinipur - 721149, West Bengal, India}
\affiliation{$^c$Saha Institute of Nuclear Physics, 1/AF Bidhannagar, Kolkata - 700064, India}
\affiliation{$^d$Homi Bhabha National Institute, Training School Complex, Anushaktinagar, Mumbai - 400085, India}

\begin{abstract}
We calculate, in a systematic way, the general structure of the self-energy of light massive fermions and the effective propagator in a thermomagnetic medium with the inclusion of anomalous magnetic moment (AMM) of the fermion in the weak field approximation. It is found that the self-energy of a massive fermion in this case consists of five non-trivial structure factors in contrast to the massless case where the self-energy contains only four. We employ the real time formalism (RTF) of thermal field theory within the ambit of hard thermal loop (HTL) approximation in the evaluation of the structure factors. The collective modes are obtained from the poles of the effective propagator of the fermion. The investigation of the dispersion relations for non-degenerate ground state shows that the effect of the magnetic field is more for up quark than the down quark because of the larger charge of the former. The important observation is that in the first excited state the degeneracy, which exists for non-zero magnetic field is lifted due to the inclusion of the AMM. It is also observed that the first excited state becomes less dispersive compared to the case when AMM is not considered, whereas the second excited state becomes more dispersive when both the magnetic field and the AMM are non-zero in comparison to the case with vanishing AMM. These effects are observed in both particle and hole-like excitations. Qualitatively similar behaviour is also seen in the case of down quarks.
	 
\end{abstract}

\maketitle
\allowdisplaybreaks

\section{Introduction}
The investigation of hot and/or dense matter in the presence of a magnetic field has captivated a broad spectrum of researchers from both theoretical and experimental domains in recent decades~\cite{Kharzeev:2013jha,Miransky:2015ava,Andersen:2014xxa,Friman:2011zz,Bali:2011qj,STAR:2021mii,An:2021wof,Milton:2021wku,Kharzeev:2022hqz}. Numerical estimates indicate that during non-central or asymmetric collisions of two heavy nuclei, extremely strong magnetic fields of the order of $ \sim 10^{18} $ Gauss or even larger can be generated due to the motion of receding spectators~\cite{Kharzeev:2007jp,Skokov:2009qp}. Additionally, strong magnetic fields are known to exist in various other physical environments. For instance, within the interiors of certain astrophysical objects known as magnetars~\cite{Duncan:1992hi,Thompson:1993hn}, magnetic fields on the order of $10^{15}$ Gauss can be found. Furthermore, there is a conjecture that primordial magnetic fields as powerful as $\sim 10^{23}$ Gauss might have been generated in the early universe during the electroweak phase transition driven by chiral anomaly~\cite{Vachaspati:1991nm,Campanelli:2013mea}. Given that the strength of these magnetic fields is comparable to the typical energy scale of Quantum Chromodynamics (QCD) ($eB\sim \Lambda_\text{QCD}^2$), various microscopic and bulk properties of strongly interacting matter could undergo significant modifications (see Refs.~\cite{Miransky:2015ava,Kharzeev:2013jha,Friman:2011zz} for recent reviews). Moreover, the existence of a robust background magnetic field gives rise to a multitude of intriguing physical phenomena~\cite{Kharzeev:2012ph,Kharzeev:2007tn,Chernodub:2010qx, Chernodub:2012tf}. These phenomena originate from the intricate vacuum structure of the underlying QCD, e.g. the Chiral Magnetic Effect (CME)~\cite{Fukushima:2008xe,Kharzeev:2007jp,Kharzeev:2009pj,Bali:2011qj}, 	Magnetic Catalysis (MC)~\cite{Shovkovy:2012zn,Gusynin:1994re,Gusynin:1995nb,Gusynin:1999pq}, Inverse Magnetic Catalysis (IMC)~\cite{Preis:2010cq,Preis:2012fh}, Chiral Vortical Effect (CVE), vacuum superconductivity and superfluidity~\cite{Chernodub:2011gs,Chernodub:2011mc} and others.

In the context of a hot and dense medium, such as a QED or QCD plasma, it is well-established that traditional bare perturbation theory encounters challenges due to the presence of infrared divergences. To address this problem originating because of massless particles, a reorganization of perturbation theory has been carried out, involving an expansion around a system of massive quasiparticles~\cite{Andersen:1999fw}. These quasiparticles acquire mass through thermal fluctuations. An essential aspect of this approach is the resummation of a specific class of diagrams known as hard thermal loop (HTL) resummation~\cite{Braaten:1989mz}. This resummation becomes necessary when the loop momenta are of the order of the temperature. An extensive literature can be found on the modification of the fermion dispersion relation, which refers to the poles of the fermion propagator, due to high-temperature effects in chirally invariant gauge theories like QCD or QED with massless fermions incorporating HTL approximations~\cite{Klimov:1981ka,Weldon:1982bn,Weldon:1989bg,Weldon:1989ys,Bellac:2011kqa,Strickland:2019tnd,Mustafa:2022got}. This modification arises from the interaction of a fermion with the thermal background in a plasma. It is found that in a parity-preserving gauge theory at finite temperature, the fermion dispersion relation leads to two solutions. These solutions, both characterized by positive energy, pertain to the propagation of fermionic excitations within the thermal medium, often referred to as quasi-particles. They are commonly known as the particle and hole/plasmino modes in the literature. Each point along these branches corresponds to specific energy and momentum values for a quasi-particle. Quasi-particles associated with the particle-like excitation possess helicity matching their chirality, while those linked to the hole-like excitation (plasmino) exhibits helicity opposite to their chirality. It is important to highlight that, the existence of the plasmino mode as an additional physical solution only arises when temperature effects are considered. 
The dispersion properties in case of a fermion with mass has been studied in Refs.~\cite{Pisarski:1989wb,Pisarski:1989cs,Petitgirard:1991mf,Quimbay:1995jn,Sumit:2022bor}. These studies have revealed that the collective mode experiences suppression as the mass of the fermion increases.

The magnetic fields produced during heavy-ion collisions are transient but its decay may be substantially delayed due to the presence of a high electrical conductivity of the hot and dense medium~\cite{Tuchin:2013apa,Tuchin:2015oka,Tuchin:2013ie,Gursoy:2014aka}. In spite of this, by the time the quark gluon plasma (QGP) equilibrates, the magnetic field strength becomes sufficiently weak so that an expansion in terms of the magnetic field is relevant and one can work in the weak field limit. In such case the relevant energy scale of the system is governed by the inequality: $|qB|\ll T^2$ where $q$ is the charge of the fermion, $B$ is the strength of the magnetic field and $T$ is the temperature of the system. A substantial body of work can be found in the literature that explores the properties of a hot and/or dense medium in the presence of a background magnetic field using HTL techniques. Refs.~\cite{Ayala:2014uua,Haque:2017nxq} have investigated the thermomagnetic correction to the three and four-point quark-gluon vertices in the presence of a weak magnetic field employing the HTL approximation. The  general structure of gluon self energy and collective excitations in a hot magnetized medium has been examined in Refs.~\cite{Hattori:2017xoo,Ayala:2018ina,Karmakar:2018aig}. The general structure of one loop self-energy of a fermion and effective quark propagator for a chirally invariant theory in a magnetized medium has been derived in Ref.~\cite{Das:2017vfh} based on which the pressure~\cite{Bandyopadhyay:2017cle} and chiral susceptibility~\cite{Ghosh:2021knc}  of a weakly magnetized hot and dense deconfined QCD matter has been evaluated. The soft contribution to the damping rate of a hard photon in a weakly magnetized QED medium has also been studied in Ref.~\cite{Ghosh:2019kmf}.

In all these investigations mentioned above the imaginary time formalism of finite temperature field theory with HTL approximation has been used. In this article, we will use the real time formalism (RTF) of thermal field theory to investigate the properties of a magnetized medium. To the best of our knowledge, this approach has not been explored earlier. Our objective is to examine the collective modes of a massive fermion in a medium with a weak space and time independent magnetic field in comparison to the temperature scale of the system. Moreover, we will also incorporate the finite values of the anomalous magnetic moment (AMM) of the fermion which appears due to the quantum corrections when fermions are coupled to the gauge fields~\cite{Peskin:1995ev,Schwartz:2014sze,Schwinger:1948iu}. Employing the Schwinger ansatz, the impact of quark AMM on various properties of strongly interacting matter has been extensively investigated using effective models~\cite{Fayazbakhsh:2014mca,Chaudhuri:2019lbw,Chaudhuri:2020lga,Mei:2020jzn,Xu:2020yag,Farias:2021fci,Wen:2021mgm,Kawaguchi:2022dbq,Mao:2022dqn,Chaudhuri:2022oru}. In the current work, we  evaluate the general structure of one loop self-energy of a massive fermion in a magnetized medium following Refs.~\cite{Weldon:1982bn,Pisarski:1989cs,Pisarski:1989wb,Petitgirard:1991mf,Das:2017vfh,Ghosh:2021knc}. Subsequently, we compute the explicit  expression of one loop self energy using the weak field expansion of the fermion propagator with finite values of AMM as obtained in Ref.~\cite{Mukherjee:2018ebw}. All the non-trivial structure factors are derived employing HTL techniques. Finally utilizing the Schwinger-Dyson method, the effective fermion propagator has been calculated and dispersion relation is investigated  by analysing the thermomagnetically modified pole of the propagator. The effective propagator thus obtained can also be used to evaluate the photon damping rate and real photon emission from a magnetized medium with AMM. 

The article is organised as follows. In Sec.~\ref{sec.self}, the general expression of one-loop self-energy of a massive fermion in a thermomagnetic dense medium is calculated employing RTF. Next in Sec.~\ref{sec.structure}, the general Dirac structure of the fermion self energy is analyzed in detail followed by the evaluation of complete interacting/effective propagator by solving Dyson-Schwinger equation in Sec.~\ref{sec.eff.prop}. Sec.~\ref{Sec_Evaluation_SF} along with its four subsections is devoted for the explicit evaluation of the structure factors using the HTL approximation. In Sec.~\ref{sec.result}, we have shown the numerical results and finally we summarize and conclude in Sec.~\ref{sec.sum}. A small Appendix~\ref{app} is provided to meet some calculational gaps of Sec.~\ref{sec.eff.prop}.

%~~~~~~~~~~~~~~~~~~~~~~~~~~~~~~~~~~~~~~~~~~~~~~~~~~~~~~~~~~~~~~~~~~~~~~~~~~~~~~~~~~~~~~~~~~~~~~~~~~~~~~~~~~~~~~~~~~~~~~~~~~~
\section{One-loop self energy of a fermion at finite temperature in presence of background field} \label{sec.self}
We start this section by specifying our choice of metric tensor which is given by $ g^\munu = \diag{ 1,-1,-1,-1} $. Any four vector $K  $ is decomposed into $ K^\mu = K^\mu_\parallel + K^\mu_\perp $ where $ K^\mu_{\parallel,\perp} = g^\munu_{\parallel,\perp} K_\nu  $ in which $ g^\munu_\parallel = \diag{1,0,0,-1} $ and $ g^\munu_\perp = \diag{0,-1,-1,0} $ so that $ g^\munu  = g^\munu_\parallel +g^\munu_\perp  $. Throughout this article, we will use capital letters to denote four-momentum and the corresponding small letters will denote the absolute values of the spatial component of the four vector (i.e. magnitude of three-momentum) e.g. $ K^\mu \equiv (k^0,\vec k) $ and $ k = |\vec k| $. The translationally invariant part of the propagator of a fermion with charge $ q $ (for example,  a up (down) quark will have a charge $ q  = 2/3 e $ ($ q  = -1/3 e $) where $ e>0 $ is the charge of a proton) and AMM $\kappa$ in the presence of a background magnetic field $\vec B = B \hat{z} $ (i.e. along the positive $ z $-direction) in weak field approximation up to first order in $ B $ is given by~\cite{Mukherjee:2018ebw}
\begin{eqnarray}
S_{B} \FB{P, m}&=& \dfrac{-\fb{\fsl{P} +m}}{P^2 -m^2 +i \epsilon} + (qB)\dfrac{i \gm^1\gm^2\fb{\fsl{P}_\parallel +m}}{\fb{P^2 -m^2 +i \epsilon}^2}+ (\kappa qB)\dfrac{\fb{\fsl{P} + m }i \gm^1\gm^2\fb{\fsl{P} +m}}{\fb{P^2 -m^2 +i \epsilon}^2} + \mathcal{O}(B^2) \nn \\
&=& \mF (P, m ; m_1) ~\Delta_F (P,m_1)\Big|_{m_1 = m} \label{Def_prop_B}~,
\end{eqnarray}
where, $\mF (P, m ; m_1) = \fb{\fsl{P} +m} +  (qB)i \gm^1\gm^2\fb{\fsl{P}_\parallel +m} \hat{\mA}_1 +(\kappa qB)\fb{\fsl{P} + m }i \gm^1\gm^2\fb{\fsl{P} +m} \hat{\mA}_1 + \mO(B^2)~,$
%\begin{equation}\label{Def_Foperator}
%\mF (P, m ; m_1) = \fb{\fsl{P} +m} +  (qB)i \gm^1\gm^2\fb{\fsl{P}_\parallel +m} \hat{\mA}_1 +(\kappa qB)\fb{\fsl{P} + m }i \gm^1\gm^2\fb{\fsl{P} +m} \hat{\mA}_1 + \mO(B^2)~,
%\end{equation}
with 
\begin{equation}\label{Def_Aop}
\hat{\mA}_n = (-1)^n \frac{\partial}{\partial (m^2_1)^n}~~~~~~\text{and}~~~~~ \Delta_F (P, m_1) = \dfrac{-1}{P^2-m_1^2 + i \epsilon}~.
\end{equation}
In Eq.~\eqref{Def_prop_B}, $ m $ is the mass and $ \kappa  $ is the AMM of the fermion respectively; also the propagator $S_{B} \FB{P, m}$ contains an identity matrix in the gauge group space which is suppressed for brevity. Note that $m_1  $ is a parameter which we set equal to the mass $ m $ after the calculation. Few comments about the propagator of a charged particle in a magnetic field are in order here. The propagator in this case separately depends upon the transverse and longitudinal components of the momentum implying that it is not translationally invariant in the configuration space due to the presence of a phase factor. The latter can be made unity by choosing an appropriate gauge transformation~\cite{Ayala:2014uua}.

Now, the $ 11 $-component of the fermion propagator in weak magnetic field approximation corresponding to Eq.~\eqref{Def_prop_B}  in the RTF of thermal field theory is given by~\cite{Mukherjee:2018ebw}
\begin{eqnarray}
S_B^{11}(P,m;m_1) &=& S_B(P,m,m_1) - \eta (P\cdot u) \TB{ S_B(P,m,m_1) - \gm^0 S^\dagger_B(P,m,m_1)\gm^0  } \nn \\
&=&  \mF (P, m ; m_1)\TB{\Delta_F (P,m_1) - 2 \pi i \eta (P\cdot u) ~\delta (P^2 - m_1^2)  }\Big|_{m_1 = m} \label{Def_porp_S11}~.
\end{eqnarray} 
In the above equation, $ u $ is the four-velocity of the heat bath and in the local rest frame $ u^\mu \equiv (1, \vec 0) $. Note that, here $ \hat{\mF} $ operates on both $ \Delta_F(P,m_1) $ and the argument of the delta function containing $m_1$. In Eq.~\eqref{Def_porp_S11}, $ \eta (P\cdot u) $ is  distribution like function and is given by $\eta(x) = \theta (x) f^+ (x) + \theta(-x) f^-(-x)$
%\begin{equation}
%\eta(x) = \theta (x) f^+ (x) + \theta(-x) f^-(-x)
%\label{Def_distn_LF_Fermion}
%\end{equation}
where, $\theta(x)$ is the unit step function and 
\begin{equation}
f^\pm (x) = \dfrac{1}{e^{(x\mp \mu)/T}+1}
\label{Def_distn_fermion}
\end{equation}
is the Fermi-Dirac distribution function.
\begin{figure}[h]
	\includegraphics[scale = 0.35]{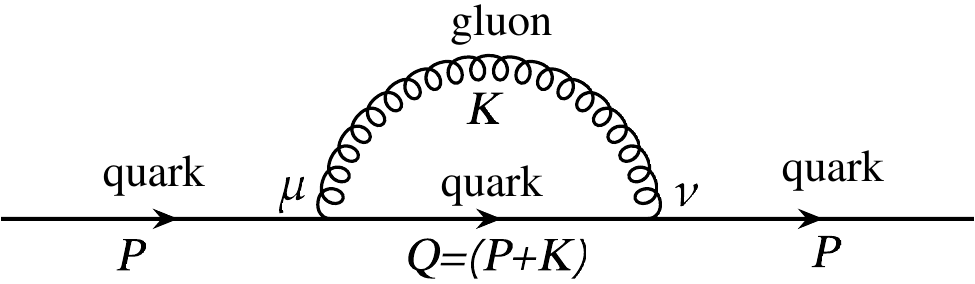}
	\caption{Typical Feynman diagram contributing to the one loop self-energy of a fermion.}
	\label{Fig_feynman}
\end{figure}

In the presence of a background magnetic field, the $ 11 $-component of one loop self-energy (see Fig.~\ref{Fig_feynman}) of the fermion can be written as 
\begin{align}
\Sigma^{11}_B (P) &= i g^2  C(R) \fourint{K} D^{11}_\munu (K)\gm^\mu S^{11}_B (Q,m,m_1) \gm^\nu \label{Def_selfE1}~,
\end{align}
where, $g$ is the gauge-fermion coupling, $ C(R) $ is the Casimir invariant of the respective gauge group and  $ D^{11}_\munu (K) $ represents the 11-component of the gluon propagator and is given by
\begin{equation}
D^{11}_\munu (K) = -g_\munu \TB{ \dfrac{-1}{K^2 + i \epsilon} + 2 \pi i N (K\cdot u ) ~\delta (K^2) }~,
\label{Def_prop_gluon}
\end{equation}
in which $N(x) = \theta(x) n(x) + \theta (-x) n(-x)$ with $n(x) = \dfrac{1}{e^{x/T}-1}$ being the Bose-Einstein distribution function.
%\begin{equation}
%N(x) = \theta(x) n(x) + \theta (-x) n(-x) ~,
%\label{Def_distn_lf_boson}
%\end{equation} 
%with 
%\begin{equation}\label{Def_distn_boson}
%n(x) = \dfrac{1}{e^{x/T}-1}~.
%\end{equation}

Now using Eqs.~\eqref{Def_porp_S11} and \eqref{Def_prop_gluon} one can simplify Eq.~\eqref{Def_selfE1} to arrive at
\begin{equation}\label{Def_selfE2}
\Sigma^{11}_B (P) = \SooV +\Soon +\Soonn ~,
\end{equation}
where,
\begin{eqnarray}
	\SooV &=&  g^2 C(R) \threeint{k} \bigg[ \hmN (k^0 = -p^0 - \wqmo) \bigg( \dfrac{1}{p^0 + \wk + \wqmo-i\epsilon }   \bigg) \nn \\ 
	&&~~~~~ -\hmN (k^0 = - \wk) \bigg( \dfrac{1}{p^0 - \wk - \wqmo+i\epsilon }   \bigg) \bigg] \dfrac{1}{4 \wk \wqmo}\bigg|_{m_1 = m}, \label{Def_S11v} \\
	\Soon &=& -g^2 C(R) \threeint{k} \bigg[ \hmN (k^ 0 = -p^0 +\wqmo) f^+ (\wqmo) \bigg(  \dfrac{1}{p^0 - \wk -\wqmo + i \epsilon}  - \dfrac{1}{p^0 + \wk -\wqmo - i \epsilon}  \bigg)  \nn \\ 
	&&~~~~~  +\hmN (k^ 0 = -p^0 -\wqmo) f^- (\wqmo) \bigg( \dfrac{1}{p^0 - \wk +\wqmo + i \epsilon}  - \dfrac{1}{p^0 +\wk +\wqmo - i \epsilon}  \bigg)  \nn \\ 
	&&~~~~~ -\hmN (k^ 0 = \wk) n(\wk) \bigg( \dfrac{1}{p^0 + \wk -\wqmo + i \epsilon}  - \dfrac{1}{p^0 +\wk +\wqmo - i \epsilon}  \bigg) \nn \\ 
	&&~~~~~ -\hmN (k^ 0 = -\wk) n (\wk) \bigg( \dfrac{1}{p^0 - \wk -\wqmo + i \epsilon}  - \dfrac{1}{p^0 -\wk +\wqmo - i \epsilon}  \bigg)  \bigg] \dfrac{1}{4 \wk \wqmo}\bigg|_{m_1 = m}, 
	\label{Def_S11n} \\
	\Soonn&=& -2 \pi ig^2 C(R)\threeint{k} \TB{ \hmN (k^0 = \wk ) f^+ (\wqmo) n(\wk) \delta (p^0 + \wk - \wqmo) \right. \nn \\ 
		&& \left.+ \hmN (k^0 = - \wk) f^- (\wqmo) n (\wk) \delta (p^0 -\wk +\wqmo)} \dfrac{1}{4 \wk \wqmo}\bigg|_{m_1 = m} \label{Def_S11n2}
\end{eqnarray}
in which $\omega_k = k$, $\omega_{q,1} = \sqrt{q^2 + m_1^2} = \sqrt{\fb{\vec{p} +\vec{k}}^2 +m_1^2}$ and 
\begin{align}
\hmN (Q) &= \gm^\mu \hmF(Q, m; m_1) \gm_\mu =  4 m -2 \fsl{Q} - i \gm^1 \gm^2 \FB{2 qB + 4m \kappa q B} \fsl{Q}_{\parallel} \hmA_1.
\label{Def_N_operator}
\end{align}
Note that Eq.~\eqref{Def_N_operator} contains derivative with respect to the mass parameter $ m_1 $ and the non-trivial Dirac structure of the self-energy. While evaluating Eq.~\eqref{Def_S11n} we have used the following identities: 
\begin{align}
\eta(Q\cdot u) &= \dfrac{1}{2 \wqmo} \TB{f^+ (\wqmo ) \delta (q^0 - \wqmo) + f^- (\wqmo) \delta (q^0 + \wqmo)} ~,\\
N (K\cdot u) &= \dfrac{1}{2 \wk} \TB{n (\wk ) \delta (k^0 - \wk) + n (\wk) \delta (k^0 + \wk)}~.
\end{align}
Since we aim to study the dispersive properties of the fermion, the relavant contribution will come from the real part of the self-energy. Now from the Eq.~\eqref{Def_S11n2}, one can observe that $ \Soonn  $ being purely imaginary will not contribute to the dispersion relation of the fermion. Moreover, in the RTF of thermal field theory the real part of the self-energy $\Sigma_B$ is equal to real part of its 11-component i.e. $ \RE \Sigma_B(P) = \RE\Soo_B(P) $~\cite{Weldon:1982bn,Mallik:2016anp,Bellac:2011kqa}. Thus simplifying Eqs.~\eqref{Def_S11v} and \eqref{Def_S11n}, we obtain
\begin{eqnarray} 
\RE~{\Sigma_B} &=&  g^2 C(R) \threeint{k} \bigg[ \hmN (k^0 = -p^0 - \wqmo) \mP \bigg( \dfrac{1}{p^0 + \wk + \wqmo } \bigg)  
	-\hmN (k^0 = - \wk) \mP \bigg( \dfrac{1}{p^0 - \wk - \wqmo }  \bigg) \nn \\
&& - \hmN (k^ 0 = -p^0 +\wqmo) f^+ (\wqmo)\mP \bigg( \dfrac{1}{p^0 - \wk -\wqmo }  - \dfrac{1}{p^0 + \wk -\wqmo } \bigg)  \nn \\ && 
-\hmN (k^ 0 = -p^0 -\wqmo) f^- (\wqmo)\mP \bigg( \dfrac{1}{p^0 - \wk +\wqmo }  - \dfrac{1}{p^0 +\wk +\wqmo } \bigg) \nn \\ && 
+\hmN (k^ 0 = \wk)n(\wk)\mP \bigg( \dfrac{1}{p^0 + \wk -\wqmo}  - \dfrac{1}{p^0 +\wk +\wqmo} \bigg) \nn \\ && 
 + \hmN (k^ 0 = -\wk) n(\wk)\mP \bigg( \dfrac{1}{p^0 - \wk -\wqmo }  - \dfrac{1}{p^0 -\wk +\wqmo } \bigg)  \bigg] \dfrac{1}{4 \wk \wqmo}\bigg|_{m_1 = m} \label{Def_ReS}
\end{eqnarray}
where $\mP$ denotes Cauchy principal value integration. In the next section we will evaluate the general structure of the fermion self-energy where we will use Eq.~\eqref{Def_ReS} to calculate the non-trivial structure factors.

%~~~~~~~~~~~~~~~~~~~~~~~~~~~~~~~~~~~~~~~~~~~~~~~~~~~~~~~~~~~~~~~~~~~~~~~~~~~~~~~~~~~~~~~~~~~~~~~~~~~~~~~~~~~~~
\section{General structure of fermion self-energy in a magnetized medium}\label{sec.structure}
In this section we discuss the general structure of the self-energy of a massive fermion in a hot magnetized medium following Refs.~\cite{Das:2017vfh,Ghosh:2021knc,Petitgirard:1991mf,Pisarski:1989cs,Pisarski:1989wb,Haque:2018eph}. Note that the fermion self-energy $\Sigma_B$ in Eq.~\eqref{Def_ReS} is a $ 4\times 4 $ matrix in the Dirac space and it is a Lorentz scalar. At finite temperature and in the presence of a background magnetic field, $\Sigma_B(P)$ is a function of four-momentum $ P $ of the corresponding fermion, medium four-velocity $ u $ and an additional four-vector $ n^\mu  $ corresponding to the magnetic field direction. It is well known that any $ 4\times 4 $ matrix can be expressed in terms of sixteen basis matrices: $\SB{\unit, \gm_5,\gm^\mu, \gm^\mu\gm_5,\sigma^{\munu}} $ where $ \gm_5  = i\gm^0\gm^1\gm^2\gm^3$ and $ \sigma^\munu = \frac{i}{2}\TB{\gm^\mu,\gm^\nu} $. Now, as argued in Ref.~\cite{Das:2017vfh}, terms involving $ \sigma^\munu $ will not arise due to its anti-symmetric nature in any loop order in the self-energy. Moreover, any term proportional to $ \gm_5 $ will not contribute as it will break the parity invariance. Thus, for the massive fermion we arrive at the following general covariant structure of the fermion self-energy
\begin{equation}\label{Def_GenS1}
\Sigma_B(P) = - \alpha - a \slashed{P} -b\slashed{u} -c \slashed{n} - \ap \gm_5 \slashed{P}- \bp \gm_5 \slashed{u}- \cp \gm_5 \slashed{n}~.
\end{equation} 
where, $\alpha$, $a$, $b$, $c$, $\ap$, $\bp$ and $\cp$ are the seven form/structure factors.  
Multiplying Eq.~\eqref{Def_GenS1} with different basis vectors and then taking the trace, one obtains all the form factors in terms of $ \Sigma_B(P) $. They are expressed as follows:
\begin{align}
\alpha &= -\dfrac{1}{4}~\Tr{\Sigma_B}~, \label{Def_alpha}\\
a &= \dfrac{1}{4}~\dfrac{\Tr{\fsl{P}\SgB } -\fb{P\cdot u} \Tr{\slashed{u}\SgB} }{\fb{P\cdot u} - P^2} ~,\label{Def_a} \\
b &= \dfrac{1}{4}~\dfrac{-\fb{P\cdot u}\Tr{\fsl{P}\SgB } + P^2 \Tr{\slashed{u}\SgB} }{\fb{P\cdot u} - P^2} ~, \label{Def_b}\\
c &= \dfrac{1}{4}~\dfrac{\fb{P\cdot n}\Tr{\fsl{P}\SgB } - \fb{P\cdot u} \fb{P\cdot n} \Tr{\slashed{u}\SgB} + \SB{\fb{P\cdot u} - P^2} \Tr{\slashed n \SgB}}{\fb{P\cdot u} - P^2}~,\\
\ap &= \dfrac{1}{4P_\perp^2}\TB{~\Tr{\gm_5\fsl{P}\SgB } - \fb{P\cdot u} \Tr{\gm_5\slashed{u}\SgB} + \fb{P\cdot n} \Tr{\gm_5\slashed n \SgB}\textcolor{white}{\dfrac{}{}}}~,\\
\bp &= \dfrac{1}{4P_\perp^2}\TB{ -\fb{P\cdot u}\Tr{\gm_5\fsl{P}\SgB } + \big\{ \fb{P\cdot n} + P^2\big\} \Tr{\gm_5\slashed{u}\SgB} - \fb{P\cdot u} \fb{P\cdot n} \Tr{\gm_5\slashed n \SgB}}~,\\
\cp &= \dfrac{1}{4P_\perp^2}\TB{ \fb{P\cdot n}\Tr{\gm_5\fsl{P}\SgB } - \fb{P\cdot u} \fb{P\cdot n} \Tr{\gm_5\slashed{u}\SgB}   + \big\{ \fb{P\cdot u} - P^2 \big\}\Tr{\gm_5\slashed n \SgB}}~.
\end{align}
where, $ P_\perp^2 = -p_x^2-p_y^2 $. Note that the structure factor $ \alpha $ is not included in Ref.~\cite{Ghosh:2021knc}, however as argued in Refs.~\cite{Pisarski:1989cs,Pisarski:1989wb,Petitgirard:1991mf,Haque:2018eph}, it should be present in the case of a massive fermion which results in explicit breaking of chiral symmetry. Now since $ \hmN (Q) $ defined in Eq.~\eqref{Def_N_operator} involves a non-trivial Dirac structure, by direct calculation from Eq.~\eqref{Def_ReS}, one can show that the structure factors $ c $ and $ \ap  $ are exactly zero for one loop self energy in the weak magnetic field limit. Although, in this article we have considered only terms linear in $ B $ in the thermo-magnetically modified fermion propagator defined in Eq.~\eqref{Def_prop_B}, we have checked that this argument is valid upto $ \mO (B^2) $ (note that, in Ref.~\cite{Ghosh:2021knc} it was shown that $ c $ and $ \ap $ are zero in absence of finite values of AMM of the fermion upto $ \mO (B^2) $). Once $ \ap = 0  $, it can be shown that $ \bp $ and $ \cp   $ take the following simpler forms:  
\begin{align}
\bp&= \dfrac{1}{4}~ \Tr{\gm_5 \slashed u \SgB} \label{Def_bp}~, \\
\cp&= -\dfrac{1}{4}~ \Tr{\gm_5 \slashed n \SgB} \label{Def_cp}.
\end{align}

Therefore the general structure of one loop self-energy of a thermo-magnetically modified massive fermion with nonzero AMM contains five non-trivial structure factors defined in Eqs.~\eqref{Def_alpha}$ - $~\eqref{Def_b}, \eqref{Def_bp} and \eqref{Def_cp} and it is given by
\begin{equation}\label{Def_SigmaB}
\SgB(P) = -\alpha  - a \slashed{P} -b\slashed{u} - \bp \gm_5 \slashed{u}- \cp \gm_5 \slashed{n}.
\end{equation}
We will derive analytical expressions for all these structure factors in Sec.~\ref{Sec_Evaluation_SF} using hard thermal loop approximation.
 
%~~~~~~~~~~~~~~~~~~~~~~~~~~~~~~~~~~~~~~~~~~~~~~~~~~~~~~~~~~~~~~~~~~~~~~~~~~~~~~~~~~~~~~~~~~~~~~ 
 \section{Effective fermion propagator} \label{sec.eff.prop}
 Employing Dyson-Schwinger equation, the effective fermion propagator can be expressed as $\mS (P) = \dfrac{1}{\fsl{P}-m- \SgB(P)}$ so that its inverse is given by
 \begin{equation}\label{Def_inv_eff_prop}
 \mS^{-1} (P) = \fsl{P}-m - \SgB(P).
 \end{equation}
 To proceed further it is useful to introduce chiral projection operators $\mathds{P}_\pm = \dfrac{1}{2} \FB{1 \pm  \gm_5}$ where $ \mdPp $ ($ \mdPm $) is the right (left) chiral projection operator. Using these projection operators, one can rewrite Eq.~\eqref{Def_GenS1} (note that co-efiicients  $ c $ and $ \ap $ are zero)~\cite{Weldon:1982bn,Das:2017vfh} as
 \begin{equation}\label{Def_SgB_proj}
 \SgB(P) = - \alpha - \mdPp \Sigma_+ \mdPm- \mdPm \Sigma_- \mdPp
 \end{equation}
 where, $\Sigma_\pm = a \slashed P + (b \mp \bp) \slashed u  \mp \cp \slashed n$. Now using the identity $ \slashed P = \mdPp \slashed P \mdPm+  \mdPm \slashed P \mdPp $, we can write the inverse of the effective propagator $\mS^{-1}$ of Eq.~\eqref{Def_inv_eff_prop} as Ref.~\cite{Weldon:1982bn} as
 \begin{equation}\label{Def_eff_inv_proj}
 \mS^{-1} (P) = \mdPp \slashed L \mdPm + \mdPm \slashed R \mdPp - C \mdPp - C \mdPm
 \end{equation}
 where,
 \begin{align}
 C& = m -\alpha \label{HTL_C} ~,\\
 L &= (1+a) P + (b-\bp) u - \cp n \label{Def_L}~,\\
 R &= (1+a) P + (b+\bp) u + \cp n~. \label{Def_R}
 \end{align}
 Now $ \mS^{-1}(P) $ as given in Eq.~\eqref{Def_inv_eff_prop} can be inverted following Ref.~\cite{Weldon:1982bn} to arrive at the expression for effective propagator for a massive fermion in magnetized medium and is given by
 \begin{align}\label{Eq_eff_prop}
 \mS (P) &= \dfrac{1}{D} \TB{ \mdPp\FB{ L^2 \slashed R -C^2 \slashed L } \mdPm+  \mdPm\FB{ R^2 \slashed L -C^2 \slashed R } \mdPp \textcolor{white}{\frac{1}{2}} \nn \right. \\ 
 	& ~~~~ \left.  + ~C \mdPp \FB{ L\cdot R - C^2 + \frac{1}{2} \TB{\slashed L, \slashed R} }\mdPp  + C \mdPm \FB{ L\cdot R - C^2 - \frac{1}{2} \TB{\slashed L, \slashed R} }\mdPm  }~,
 \end{align} 
 where the denominator of the propagator reads
 \begin{equation}\label{Def_D}
 D = L^2 R^2 - 2 C^2 \FB{L \cdot R} + C^4~.
 \end{equation}
The denominator $D$ can be factorized as shown in Appendix~\ref{app} and can be put in a more useful form like
\begin{align}\label{Def_D_useful}
	D =  D_+(p^0,p) ~D_-(p^0,p)~,
\end{align}
where 
\begin{align}
	D_\pm(p^0,p;eB) &=  \fb{\mtA p^0 +\mtB} \mp \sqrt{\mtC^2 - \mtD^2 + \mtA^2 p^2 }~,
\end{align}
in which the quantities $\mtA$, $\mtB$, $\mtC$ and $\mtD$ are functions of the structure factors $\alpha$, $a$, $b$, $c$, $\ap$, $\bp$ and $\cp$ whose explicit expressions are provided in Appendix~\ref{app}. The poles of the propagator in Eq.~\eqref{Eq_eff_prop} give the dispersion relations $\omega=\omega(p)$ of the fermion which essentially requires solving $D(p^0=\omega,p)=0$; or in other words solving the following set of transcendental equations for $\omega$:
\begin{align}
D_+(p^0=\omega,p) &=0 \label{HTL_Dp}~, \\
D_-(p^0=\omega,p) &=0 \label{HTL_Dm}.
\end{align}

%~~~~~~~~~~~~~~~~~~~~~~~~~~~~~~~~~~~~~~~~~~~~~~~~~~~~~~~~~~~~~~~~~~~~~~~~~~~~~~~~~~~~~ 
\section{Evaluation of the structure factors}\label{Sec_Evaluation_SF}
In this section we will derive the non-trivial structure factors appearing in Eq.~\eqref{Def_SigmaB} within the ambit of HTL approximation. We will consider the loop momentum $ K $ as hard ($ \sim T $) and the external momentum ($ \sim gT $) will be considered negligible compared to $ K $. General remarks on the HTL approximation can be found in Refs.~\cite{Bellac:2011kqa,Laine:2016hma,Strickland:2019tnd,Mustafa:2022got}. 
Note that, the medium independent part of Eq.~\eqref{Def_ReS} will renormalise the vacuum. So we need to work with only medium-dependent part.
 \subsection{Evaluation of $ \alpha $}
 Considering only the medium dependent part of Eq.~\eqref{Def_ReS} in Eq.~\eqref{Def_alpha} we get
 \begin{align}
 \alpha &= -g^2 C(R) \threeint{k} \bigg[ \dfrac{1}{4} \Tr{ \hmN (k^ 0 = -p^0 +\wqmo)} f^+ (\wqmo)\mP \bigg( \dfrac{1}{p^0 - \wk -\wqmo }  - \dfrac{1}{p^0 + \wk -\wqmo }  \bigg)  \nn \\ & 
 	\hspace{1 cm}+\dfrac{1}{4} \Tr{ \hmN (k^ 0 = -p^0 -\wqmo) }f^- (\wqmo)\mP \bigg( \dfrac{1}{p^0 - \wk +\wqmo }  - \dfrac{1}{p^0 +\wk +\wqmo } \bigg)  \nn \\ & 
 	\hspace{1 cm}-\dfrac{1}{4} \Tr{ \hmN (k^ 0 = \wk)} n(\wk)\mP \bigg( \dfrac{1}{p^0 + \wk -\wqmo}  - \dfrac{1}{p^0 +\wk +\wqmo} \bigg)  \nn \\ & 
 	\hspace{1 cm}-\dfrac{1}{4} \Tr{ \hmN (k^ 0 = -\wk)} n(\wk)\mP \bigg( \dfrac{1}{p^0 - \wk -\wqmo }  - \dfrac{1}{p^0 -\wk +\wqmo } \bigg)  \bigg] \dfrac{1}{4 \wk \wqmo}\bigg|_{m_1 = m} \nn \\
 %
%&= 4 m g^2 C(R) \threeint{k} \bigg[   f^+ (\wq)\mP \bigg( \dfrac{1}{p^0 - \wk -\wq }  - \dfrac{1}{p^0 + \wk -\wq } \bigg)  \nn \\ & 
% 	\hspace{1 cm}+f^- (\wq)\mP \bigg( \dfrac{1}{p^0 - \wk +\wq }  - \dfrac{1}{p^0 +\wk +\wq }  \bigg)- n(\wk)\mP \bigg( \dfrac{1}{p^0 + \wk -\wq}  - \dfrac{1}{p^0 +\wk +\wq} \bigg)  \nn \\ & 
% 	\hspace{1 cm}- n(\wk)\mP \bigg( \dfrac{1}{p^0 - \wk -\wq }  - \dfrac{1}{p^0 -\wk +\wq } \bigg)  \bigg] \dfrac{1}{4 \wk \wq} \label{Def_Deri_al1}~,
%
&= 4 m g^2 C(R) \threeint{k} \bigg[   f^+ (\wq)\mP \bigg( \dfrac{1}{p^0 - \wk -\wq }  - \dfrac{1}{p^0 + \wk -\wq } \bigg)   
+f^- (\wq)\mP \bigg( \dfrac{1}{p^0 - \wk +\wq }  - \dfrac{1}{p^0 +\wk +\wq }  \bigg) 
\nn \\ &- n(\wk)\mP \bigg( \dfrac{1}{p^0 + \wk -\wq}  - \dfrac{1}{p^0 +\wk +\wq} \bigg)  
- n(\wk)\mP \bigg( \dfrac{1}{p^0 - \wk -\wq }  - \dfrac{1}{p^0 -\wk +\wq } \bigg)  \bigg] \dfrac{1}{4 \wk \wq} \label{Def_Deri_al1}~,
 \end{align}
 where we have used the fact that $ \Tr{\hmN} = -16 m $. Next we will consider the scenario where the chemical potential of the fermion is zero in which case the Fermi-Dirac distribution functions become $f^\pm(x) \to f(x) = [e^{x/T}+1]^{-1}$ (say). In the HTL approximation we can thus write
 \begin{align}
 \wq  &\simeq \wk + \vec p \cdot \hat{k} \label{Def_HTL1}~,\\
 f(\wq) &\simeq f(\wk) + \vec p \cdot \hat{k} \frac{d f(\wk)}{d \wk}~, \label{Def_HTL2}\\
 \dfrac{1}{p^0 \pm (\wk - \wq)} &\simeq \dfrac{1}{p^0 \mp \pkcap}~, \label{Def_HTL3}\\
 \dfrac{1}{p^0 \pm  (\wk + \wq)} &\simeq \dfrac{1}{\pm 2 \wk }\label{Def_HTL4}~.
 \end{align}  
 Making use of Eqs~\eqref{Def_HTL1}-\eqref{Def_HTL4} in Eq.~\eqref{Def_Deri_al1}, we obtain
  \begin{align}
 	\alpha &\simeq 4 m g^2 C(R) \threeint{k} \frac{1}{4 \wk \wq} \bigg[ \SB{f(\wk)+ n(\wk)} \bigg( \dfrac{1}{ p^0 + \pkcap }  - \dfrac{1}{p^0 - \pkcap}  \bigg)  + \dfrac{d f(\wk)}{d \wk} \bigg(  \dfrac{\pkcap}{ p^0 + \pkcap }  - \dfrac{\pkcap}{p^0 - \pkcap}  \bigg)   \bigg] \nn \\
 	&\simeq - 8 m g^2 C(R) \threeint{k} \dfrac{1}{4 \wk \wq} \dfrac{d f(\wk)}{d \wk} \dfrac{\pkcap}{p^0 -\pkcap}  \nn 
 	\\
 	&\simeq - \dfrac{8 m g^2 C(R)}{8 \pi^2} \ZtoInfint d k  \dfrac{d f(\wk)}{d \wk} \AngInt \dfrac{\pkcap}{p^0 - \pkcap}~.\label{Def_Deri_al2}
 \end{align} 
 Here to arrive at the second line we have changed the integration variable from $ \vec k \to -\vec k$ in  all the terms that contain $ p^0 +\pkcap $ in the denominator and observed that only the last term survives.  Defining a four vector $ \hat{K}^\mu = (1, \hat{k}) $ such that $ \PKcap = p^0 - \pkcap$, the integral with respect to $ k $ can be evaluated exactly and we get in the HTL approximation
 \begin{equation}\label{HTL_alpha}
 \alpha \simeq \dfrac{8 m g^2 C(R)}{ 16 \pi^2} \AngInt \dfrac{\pkcap}{\PKcap}~.
 \end{equation}

 %~~~~~~~~~~~~~~~~~~~~~~~~~~~~~~~~~~~~~~~~~~~~~~~~~~~~~~~~~~~~~~~~~~~~~~~~~~~~~~~
 \subsection{Evaluations of $ a $ and $ b $}
 To calculate the coefficients $ a $ and $ b  $ defined in Eqs.~\eqref{Def_a} and \eqref{Def_b} respectively, one needs to use the following values of the traces:
 \begin{align}
 \Tr{ \slashed P \hmN (Q)} &= 8 P \cdot Q \label{eq.tr.1}\\
 \Tr{\slashed u \hmN (Q)} &=  - 8 Q\cdot u~. \label{eq.tr.2}
 \end{align} 
 Thus, it is evident from Eqs.~\eqref{eq.tr.1} and \eqref{eq.tr.2}, the coefficients $ a $ and $ b $ do not receive any correction due to the presence of background magnetic field. So, in the HTL approximation, they will have the same expression as obtained in Refs.~\cite{Weldon:1982bn,Bellac:2011kqa,Laine:2016hma,Mustafa:2022got} 
 \begin{align}
 a &\simeq -\frac{\Mth^2}{p^2} \AngInt \frac{\pkcap}{\PKcap} ~,\label{eq.a}\\
 b &\simeq \frac{\Mth^2}{p^2} \AngInt \frac{(P\cdot u) (\pkcap) -p^2}{\PKcap}~ \label{eq.b}
 \end{align} 
where $ \Mth^2 = g^2 C(R)T^2/8 $. Note that, the background field independence of the structure factors $ a $ and $ b $ is a manifestation of the fact that we are working up to first order in $ B  $. However, if one goes to higher order in $ B $, the expressions for $ a $ and $ b $ will depend on the magnetic field~\cite{Ghosh:2021knc}.
 
 %~~~~~~~~~~~~~~~~~~~~~~~~~~~~~~~~~~~~~~~~~~~~~~~~~~~~~~~~~~~~~~~~~~~~~~~~~~~~~~~~~~~~~~~~~~~~~~~~~~~~~~~~~~~
 \subsection{Evaluation of $ \bp $}
To evaluate $\bp $, we again use the temperature dependent part of Eq.~\eqref{Def_ReS} in Eq.~\eqref{Def_bp} to arrive at
\begin{align}
\bp &= -g^2 C(R) \threeint{k} \bigg[ \dfrac{1}{4} \Tr{\gm_5 \slashed{u} \hmN (k^ 0 = -p^0 +\wqmo)} f^+ (\wqmo)\mP \bigg( \dfrac{1}{p^0 - \wk -\wqmo }  - \dfrac{1}{p^0 + \wk -\wqmo }  \bigg)  \nn \\ & 
	\hspace{1 cm}+\dfrac{1}{4} \Tr{\gm_5 \slashed{u} \hmN (k^ 0 = -p^0 -\wqmo) }f^- (\wqmo)\mP \bigg( \dfrac{1}{p^0 - \wk +\wqmo }  - \dfrac{1}{p^0 +\wk +\wqmo } \bigg) \nn \\ & 
	\hspace{1 cm}-\dfrac{1}{4} \Tr{\gm_5 \slashed{u} \hmN (k^ 0 = \wk)} n(\wk)\mP \bigg( \dfrac{1}{p^0 + \wk -\wqmo}  - \dfrac{1}{p^0 +\wk +\wqmo} \bigg)  \nn \\ & 
	\hspace{1 cm}-\dfrac{1}{4} \Tr{ \gm_5 \slashed{u}\hmN (k^ 0 = -\wk)} n(\wk)\mP \bigg( \dfrac{1}{p^0 - \wk -\wqmo }  - \dfrac{1}{p^0 -\wk +\wqmo } \bigg)  \bigg] \dfrac{1}{4 \wk \wqmo}\bigg|_{m_1 = m} \nn \\
&= 4 m g^2 C(R) \threeint{k} \fb{ 2 qB  + 4 \kappa qB m} q_z \hmA_1 \bigg[  f^+ (\wqmo)\mP \bigg( \dfrac{1}{p^0 - \wk -\wqmo }  - \dfrac{1}{p^0 + \wk -\wqmo } \bigg) \nn \\ & 
	+f^- (\wqmo)\mP \bigg( \dfrac{1}{p^0 - \wk +\wqmo }  - \dfrac{1}{p^0 +\wk +\wqmo } \bigg)- n(\wk)\mP \bigg( \dfrac{1}{p^0 + \wk -\wqmo}  - \dfrac{1}{p^0 +\wk +\wqmo} \bigg)  \nn \\ & 
	- n(\wk)\mP \bigg( \dfrac{1}{p^0 - \wk -\wqmo }  - \dfrac{1}{p^0 -\wk +\wqmo } \bigg) \bigg] \dfrac{1}{4 \wk \wqmo} \bigg|_{m_1 = m}  \label{Def_Deri_bp1}~.
\end{align} 
 
 Notice that $ \Tr{\gm_5 \slashed{u} \hmN } $  receives contributions only from the background field dependent part of the one loop self-energy which is evident from the expression of $ \hmN $ given in Eq.~\eqref{Def_N_operator}. Thus, the structure factor $ \bp  $  arises purely due to the presence of background magnetic field.  
 Here also we will work with a system at zero chemical potential. Furthermore, we will assume that the Bose-Einstein distribution depends on the quark mass i.e.  
 $\wk \to \omega_{k,1} = \sqrt{k^2 + m_1^2 } $. This non-zero quark mass works as an infrared regulator that allows one to calculate the leading temperature behaviour~\cite{Ayala:2014uua}. Moreover, in the HTL approximation we can write $ q_z \simeq k_z $. Incorporating these along with the general HTL results expressed in Eqs.~\eqref{Def_HTL1}$ - $\eqref{Def_HTL4} we get
 \begin{align}
 \bp &\simeq \FB{2 qB + f \kappa qB m} g^2 C(R) \ParOne{ }{ m_1^2} \threeint{k} \bigg[  \SB{f(\omega_{k,1})+n (\omega_{k,1})} \bigg(\dfrac{k_z}{p^0 + \pkcap }  - \dfrac{k_z}{p^0 -\pkcap }  \bigg) \nn \\ 
 & \hspace{6cm}  + \dfrac{d f(\wk)}{d \wk} \bigg\{  \dfrac{(\pkcap) k_z}{ p^0 + \pkcap }  - \dfrac{(\pkcap) k_z}{p^0 - \pkcap}  \bigg\}   \bigg] \dfrac{1}{4 \wk \omega_{k,1}}\bigg|_{m_1 = m}  \nn \\
 &\simeq - (4 qB + 8 \kappa qB m) g^2 C(R) \ParOne{}{m_1^2} \threeint{k} \frac{1}{4 \wk \omega_{k,1}} \SB{ f(\omega_{k,1})  + n(\omega_{k,1}) } \frac{k_z}{p^0 -\pkcap}\bigg|_{m_1 = m} \nn \\
 &\simeq \dfrac{(4 qB + 8 \kappa qB m)}{8 \pi^2} \ParOne{}{m_1^2} \ZtoInfint \dfrac{k^2 d k}{\omega_{k,1} } \SB{ f(\omega_{k,1})  + n(\omega_{k,1})  } \AngInt \frac{\Kcapn}{\PKcap}\bigg|_{m_1 = m}  ~.\label{Def_Deri_bp2}
 \end{align}
 Here, in the first line we have neglected terms subleading in $ T $ appearing due to derivative with respect to $ m_1^2 $ which will contain higher powers of $ p^0\pm \wk \pm \wqmo $ in the denominator~\cite{Das:2017vfh,Ayala:2014uua}. To arrive at the third line we use the same trick as done while deriving $\alpha $ and change the variable from $ \vec k \to -\vec k $ in all the terms containing $ p^0-\pkcap $ in the denominator. Here only the first term within square bracket is nonzero. In the fourth line we have used  $ \Kcapn  = -\cos \theta_k $, where $\theta_k $   is the angle between $ \vec k $ and $ z $-axis. 
 Defining new variables $ x = k/T $ and $ y = m_1/T $, Eq.~\eqref{Def_Deri_bp2} can be rewritten as
 \begin{align}
 \bp &\simeq  \dfrac{qB + 2 \kappa qB m}{2 \pi^2} \ParOne{}{y^2}\ZtoInfint\dfrac{dx x^2}{\sqrt{x^2+y^2}  } \SB{ f(T\sqrt{x^2 +y^2}) + n(T\sqrt{x^2 +y^2}) } \AngInt \dfrac{\Kcapn}{\PKcap} \bigg|_{m_1 = m} 
 \end{align}
 which can now be expressed in terms of the following standard integrals~\cite{Dolan:1973qd,Kapusta:2006pm,Ayala:2014uua}
 \begin{align}
 h_n^\pm(y) &= \dfrac{1}{\Gamma (n) } \ZtoInfint\dfrac{dx x^{n-1}}{\sqrt{x^2+y^2}  }  \dfrac{1}{e^{\sqrt{x^2+y^2}}  \mp1}\label{Def_Int_id1}~,
% \\ f_n(y) &= \dfrac{1}{\Gamma (n) } \ZtoInfint\dfrac{dx x^{n-1}}{\sqrt{x^2+y^2}  }  \dfrac{1}{e^{\sqrt{x^2+y^2}} +1}\label{Def_Int_id2}~,
 \end{align}
 as 
 \begin{align}
 \bp &\simeq   -  \dfrac{qB + 2 \kappa qB m}{4 \pi^2} \SB{ f_1(y) + h_1(y) } \AngInt \dfrac{\Kcapn}{\PKcap}\bigg|_{m_1 = m} 
 \end{align}
 where we have used the relations $\ParOne{h^\pm_{n+1}}{y^2} = -\dfrac{h^\pm_{n-1}}{2 n}$ satisfied by the functions $ h^\pm_n(y)$.
 Now using the high temperature expansion for $ h^\pm_1(y) $ given in Ref.~\cite{Dolan:1973qd,Kapusta:2006pm,Ayala:2014uua}, we arrive at the final expression for the structure factor $ \bp $ in the HTL approximation as
 \begin{equation}
 \bp \simeq 4 g^2 C(R) M^2(qB, \kappa ;T) \AngInt \dfrac{\Kcapn}{\PKcap} \label{eq.bp}
 \end{equation}
  where the magnetic mass is given by
  \begin{equation}\label{HTL_M2}
  M^2(qB, \kappa;T )  = \dfrac{qB + 2 \kappa qB m}{16 \pi^2} \FB{\ln 2 - \frac{\pi T}{2 m}}~.
  \end{equation}
  Note that, the above expression is different from the previously obtained  magnetic mass in Ref.~\cite{Ayala:2014uua,Haque:2017nxq,Das:2017vfh} due to the presence of AMM.

 %~~~~~~~~~~~~~~~~~~~~~~~~~~~~~~~~~~~~~~~~~~~~~~~~~~~~~~~~~~~~~~~~~~~~~~~~~~~~~~~~~~~~~~~~~~~~~~~~~~~~~~~~ 
 \subsection{Evaluation of $ \cp $}
 Again from Eq.~\eqref{Def_cp} we get
 \begin{align}
 \cp &= -g^2 C(R) \threeint{k} \bigg[ \dfrac{1}{4} \Tr{\gm_5 \slashed{n} \hmN (k^ 0 = -p^0 +\wqmo)} f^+ (\wqmo)\mP \bigg( \dfrac{1}{p^0 - \wk -\wqmo }  - \dfrac{1}{p^0 + \wk -\wqmo } \bigg)  \nn \\ 
 & 	\hspace{1 cm}+\dfrac{1}{4} \Tr{\gm_5 \slashed{n} \hmN (k^ 0 = -p^0 -\wqmo) }f^- (\wqmo)\mP \bigg( \dfrac{1}{p^0 - \wk +\wqmo }  - \dfrac{1}{p^0 +\wk +\wqmo } \bigg) \nn \\ 
 & 	\hspace{1 cm}-\dfrac{1}{4} \Tr{\gm_5 \slashed{n} \hmN (k^ 0 = \wk)} n(\wk)\mP \bigg( \dfrac{1}{p^0 + \wk -\wqmo}  - \dfrac{1}{p^0 +\wk +\wqmo} \bigg)  \nn \\ 
 & 	\hspace{1 cm}-\dfrac{1}{4} \Tr{ \gm_5 \slashed{n}\hmN (k^ 0 = -\wk)} n(\wk)\mP \bigg( \dfrac{1}{p^0 - \wk -\wqmo }  - \dfrac{1}{p^0 -\wk +\wqmo } \bigg)  \bigg] \dfrac{1}{4 \wk \wqmo}\bigg|_{m_1 = m} \nn \\
 &= 4 m g^2 C(R) \threeint{k} \fb{ 2 qB  + 4 \kappa qB m}  \hmA_1 \bigg[  \wqmo~  f^+ (\wqmo)\mP \bigg( \dfrac{1}{p^0 - \wk -\wqmo }  - \dfrac{1}{p^0 + \wk -\wqmo } \bigg)  \nn \\ 
 & \hspace{.5 cm} - \wqmo~  f^- (\wqmo)\mP \bigg( \dfrac{1}{p^0 - \wk +\wqmo }  - \dfrac{1}{p^0 +\wk +\wqmo } \bigg) -(p^0 + \wk) n(\wk)\mP \bigg(\dfrac{1}{p^0 + \wk -\wqmo}  - \dfrac{1}{p^0 +\wk +\wqmo} \bigg)  \nn \\ 
 & \hspace{1 cm} - (p^0  - \wk)n(\wk)\mP \bigg( \dfrac{1}{p^0 - \wk -\wqmo }  - \dfrac{1}{p^0 -\wk +\wqmo } \bigg)  \bigg] \dfrac{1}{4 \wk \wqmo} \bigg|_{m_1 = m}  \label{Def_Deri_cp1}~.
 \end{align} 
 Here we have used the fact that $ \Tr{\gm_5 \slashed n \hmN } = 2q^0(2 qB + 4 \kappa q B m) =  2(k^0 + p^0)(2 qB + 4 \kappa q B m) $. 
 Using similar steps as described after Eq.~\eqref{Def_Deri_bp1}, one can show that in the HTL approximation
 \begin{equation}
 \cp \simeq - 4g^2 C(R) M^2 (qB, \kappa ; T) \AngInt \frac{\Kcapu}{\PKcap}~. \label{eq.cp}
 \end{equation}
 Notice that $ \cp $ will be finite only in a magnetized medium.

 Finally, we note that all the expressions of the structure factors obtained in Eqs.~\eqref{HTL_alpha}, \eqref{eq.a}, \eqref{eq.b}, \eqref{eq.bp} and \eqref{eq.cp} contain angular integrals yet to be evaluated. Following Refs.~\cite{Das:2017vfh,Weldon:1982bn,Mustafa:2022got,Bellac:2011kqa,Laine:2016hma}, one can explicitly perform these integrals. Here we note the final results:
 \begin{align}
 \alpha &\simeq -4 g^2 C(R) \frac{2m}{16 \pi^2} Q_1 (p^0/p)~,\\
 a &\simeq -\dfrac{\Mth^2}{p^2} Q_1 \fb{p^0/p} ~,\\
 b &\simeq \frac{\Mth^2}{p} \TB{ \frac{p^0}{p} Q_1 (p^0/p)  - Q_0 (p^0/p)}~,\\
 \bp &\simeq - 4 g^2 C(R) M^2 (qB, \kappa;T) \frac{p_z}{p^2} Q_1 (p^0/p) ~,\\
 \cp & \simeq - 4 g^2 C(R) M^2 (qB, \kappa;T) \frac{1}{p} Q_0 (p^0/p) ~,
 \end{align}
where $Q_0(x) = \frac{1}{2}\ln\FB{\frac{1+x}{1-x}}$ and $Q_1(x) = xQ_0(x)-1$ are the Legendre functions of second kind. One should note that all these structure factors for one loop self energy of a massive fermion in a magnetized medium derived in the weak field approximation  are valid in the range $ \Mth^2 (\sim g^2T^2) \ll |q_f eB| \ll T^2$~\cite{Ayala:2014uua,Das:2017vfh}.

 %~~~~~~~~~~~~~~~~~~~~~~~~~~~~~~~~~~~~~~~~~~~~~~~~~~~~~~~~~~~~~~~~~~~~~~~~~~~~~~~~~~~~~~~~~~~~~~~~~~~~~~~~~~~~~~~~~~~~~~~~~~
 \section{Numerical Results \& Discussions} \label{sec.result}
 In this section we present the numerical results for the collective oscillations of quarks in a hot magnetized medium with explicit breaking of chiral symmetry. Throughout this section we will consider a system of low lying quarks (i.e.  up and down quarks). The numerical values of medium temperature  and strength of the strong coupling constant will be taken as $ T = 200 $~MeV and $ \alpha_s = \frac{g^2}{4\pi} = 0.3 $~\cite{Das:2017vfh}. Moreover, the value of Casimir invariant $ C(R)$ is taken to be $C(R) = 4/3 $  for the $ SU(3) $ group. Thus with the current choice of parameters $ \Mth \sim 160 $ MeV. It should be noted that, all the numerical results are obtained using $ m = M_{\rm th} $.
 \begin{figure}[hbt]
 	\includegraphics[scale = 0.34]{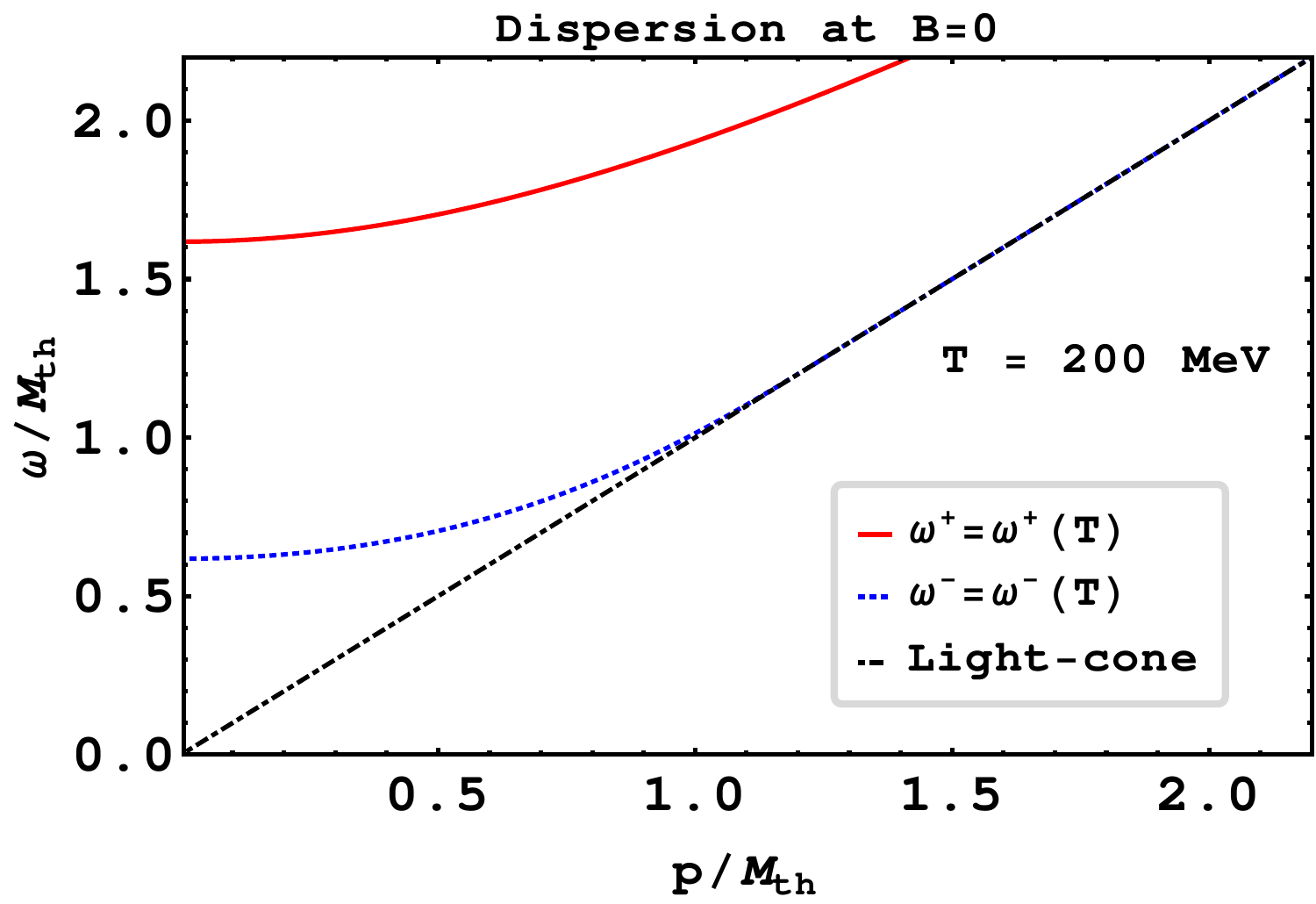}
 	\caption{Collective modes of quarks with explicit chiral symmetry breaking in an plasma. The solid red line represents the particle-like excitation, while the blue dotted line represents the hole-like excitation. }
 	\label{Fig_DeB0}
 \end{figure}
\begin{figure}[hbt]
	\includegraphics[scale=0.34]{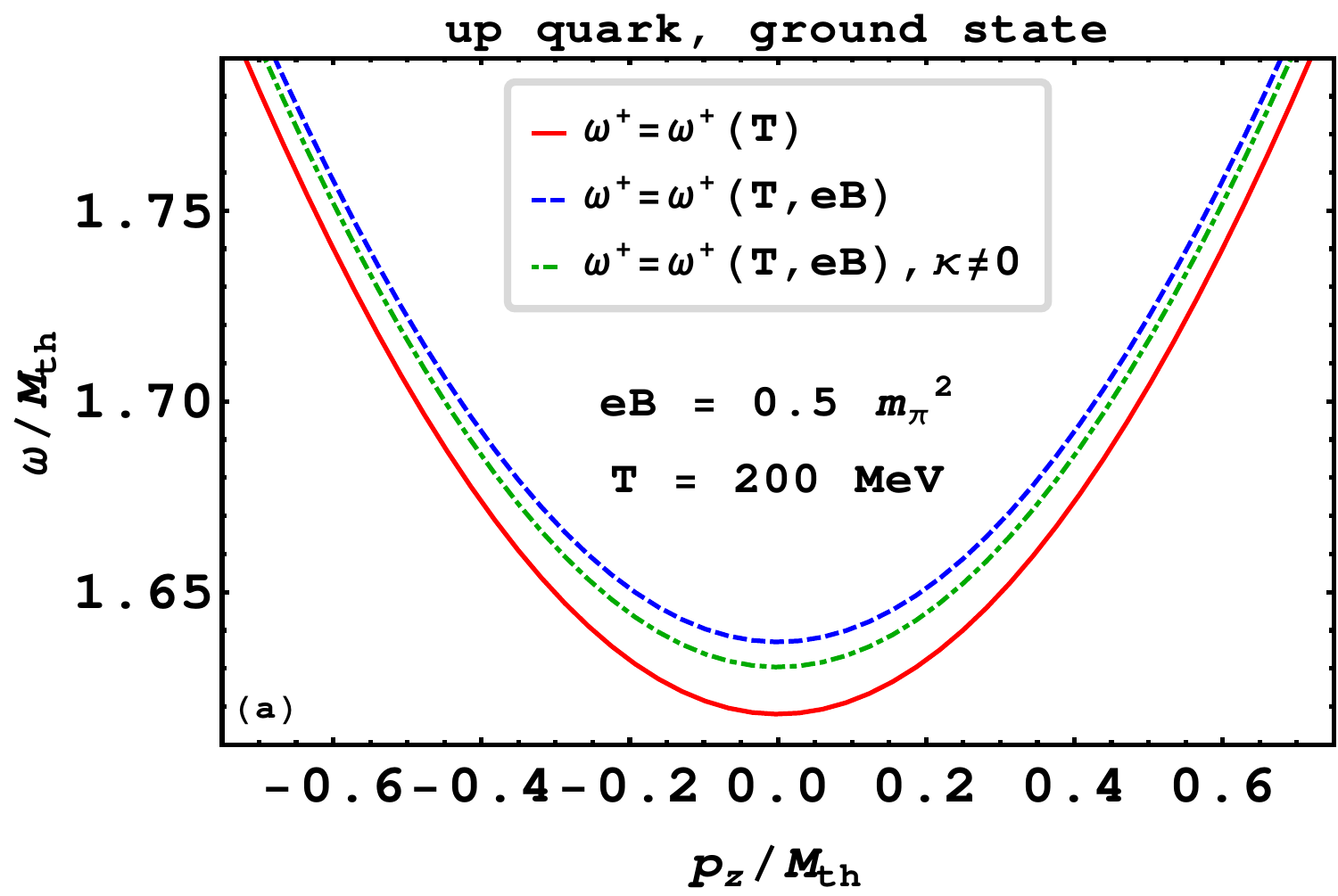} ~~ 
	\includegraphics[scale=0.34]{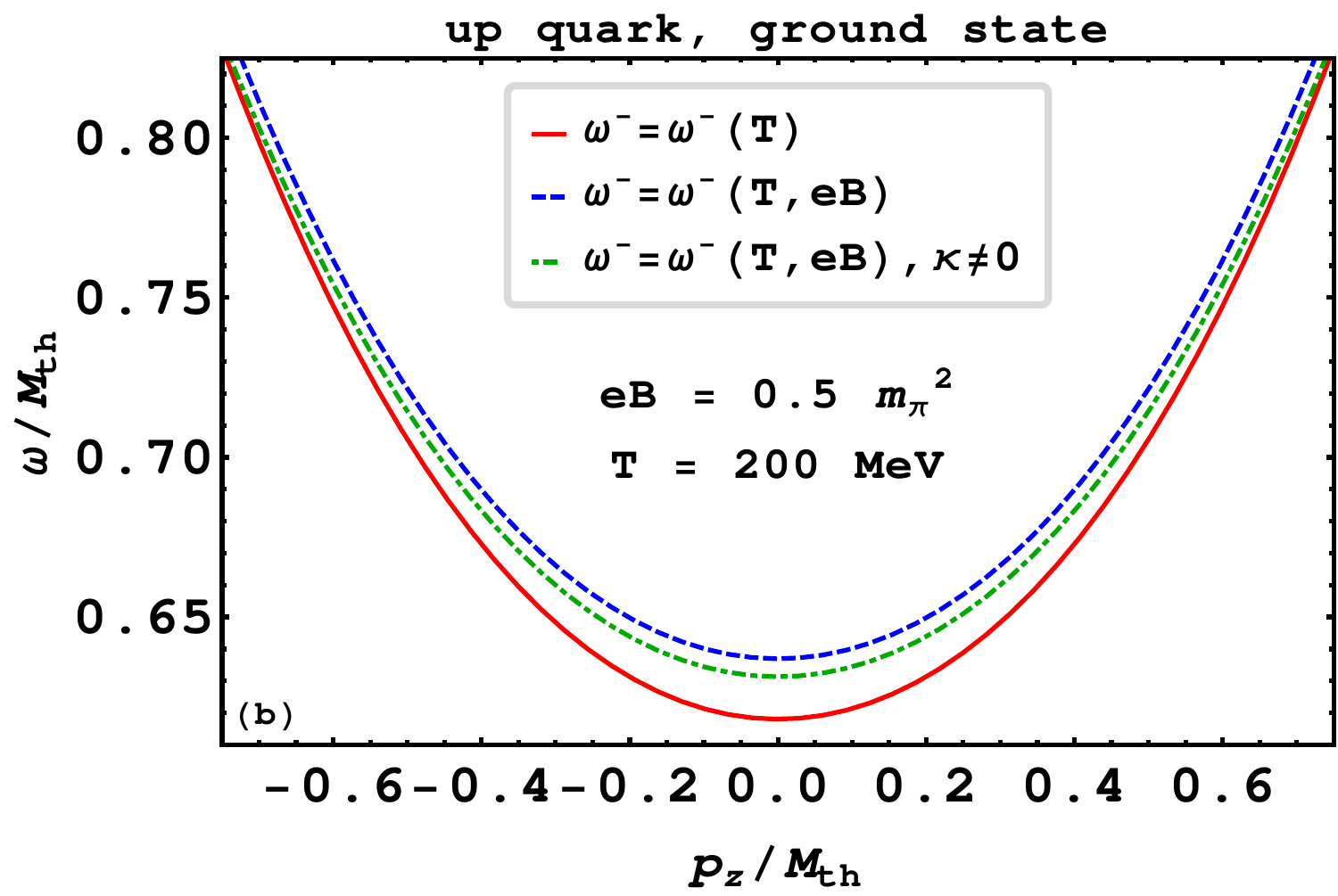}\\
	\includegraphics[scale=0.34]{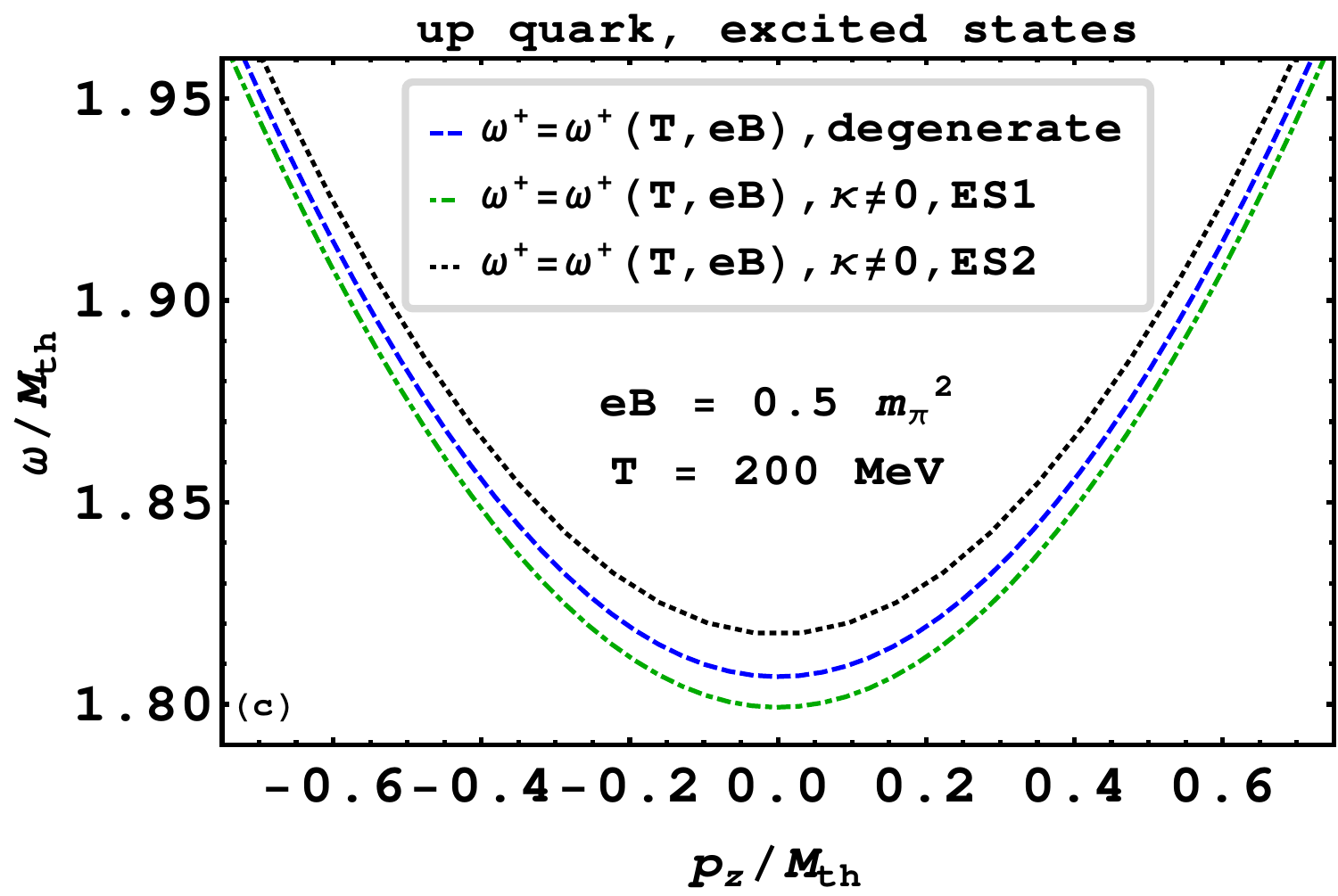} ~~ 
	\includegraphics[scale=0.34]{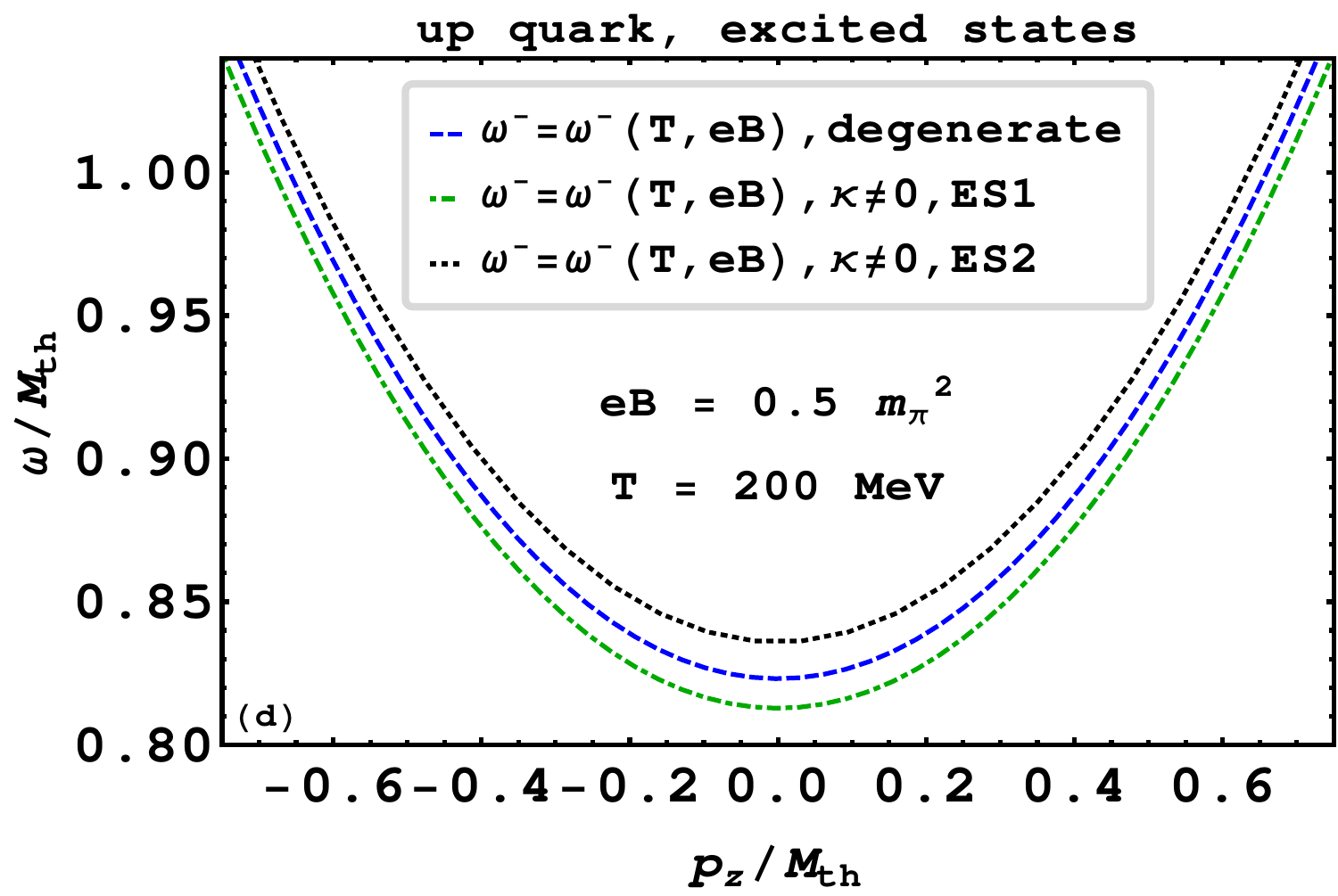}
	\caption{Dispersion of up quark in hot magnetized medium with finite values of AMM at $ T = 200 $ MeV and $ eB = 0.5~m_\pi^2 $ in the weak field limit. In the left (right) panel we have shown the particle (hole) like solutions. In the top (bottom) panel we have plotted the ground (excited) state. `ES1' and `ES2' refer to non-degenerate excited states with $\SB{n,s=1,1}$ and $\SB{n,s= 0,-1} $ respectively. Here $ \SB{n, s} $ labels the quantized values of the energy eigenvalues of Dirac equation with finite values of AMM as given in Eq.~\eqref{Eq_Dispersion}.}
	\label{Fig_Dup}
\end{figure}
\begin{figure}[hbt]
	\includegraphics[scale=0.34]{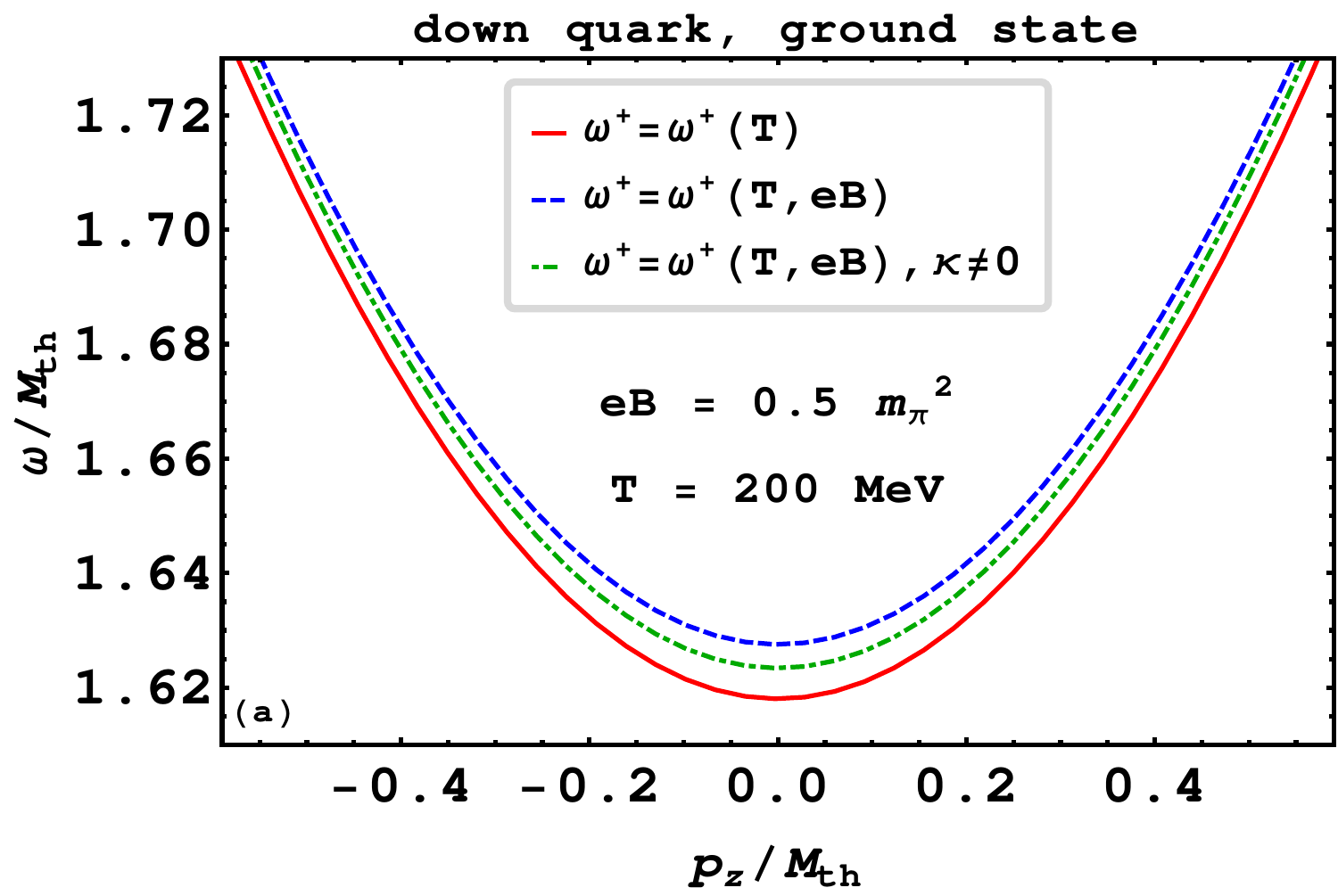} ~~ 
	\includegraphics[scale=0.34]{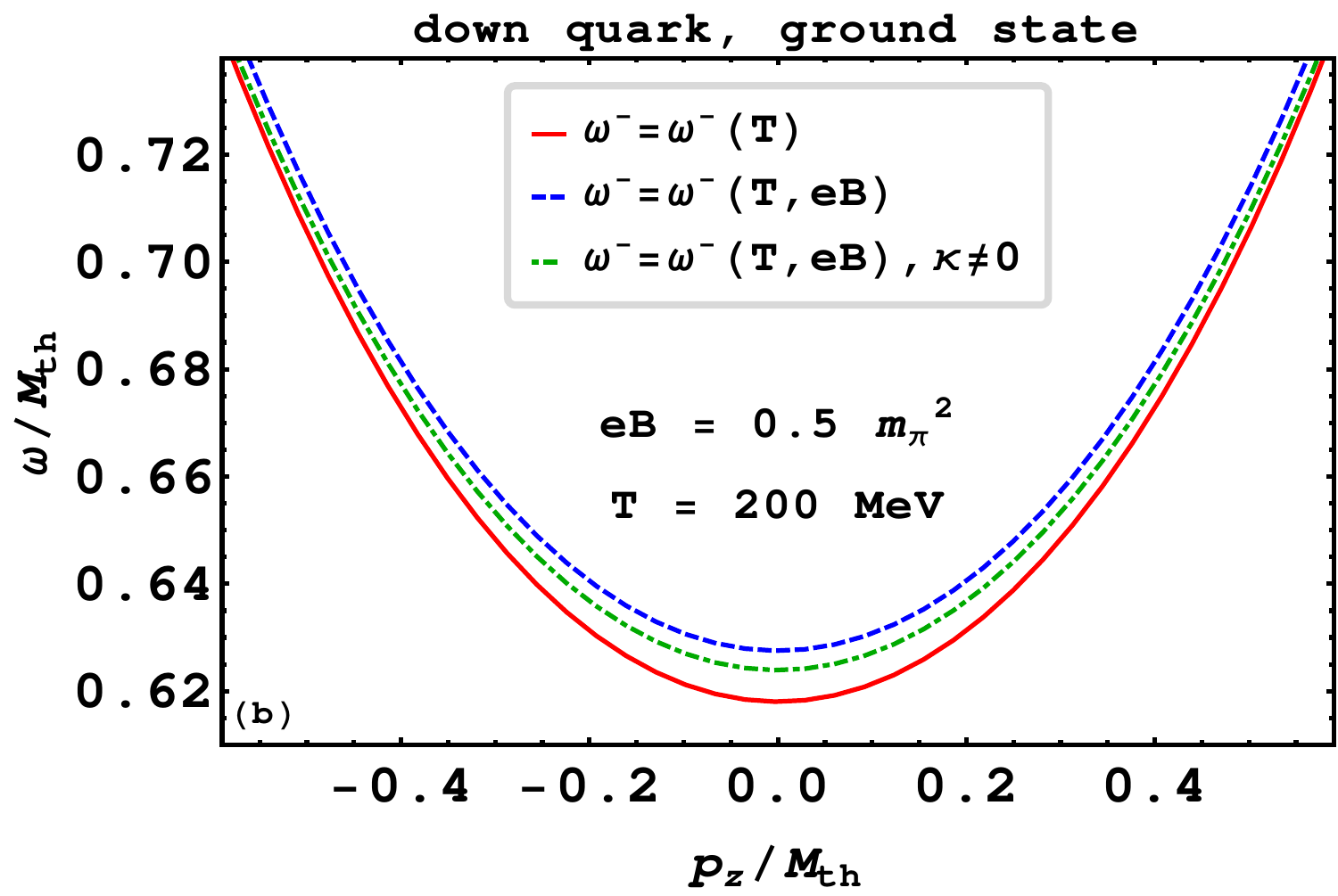}\\
	\includegraphics[scale=0.34]{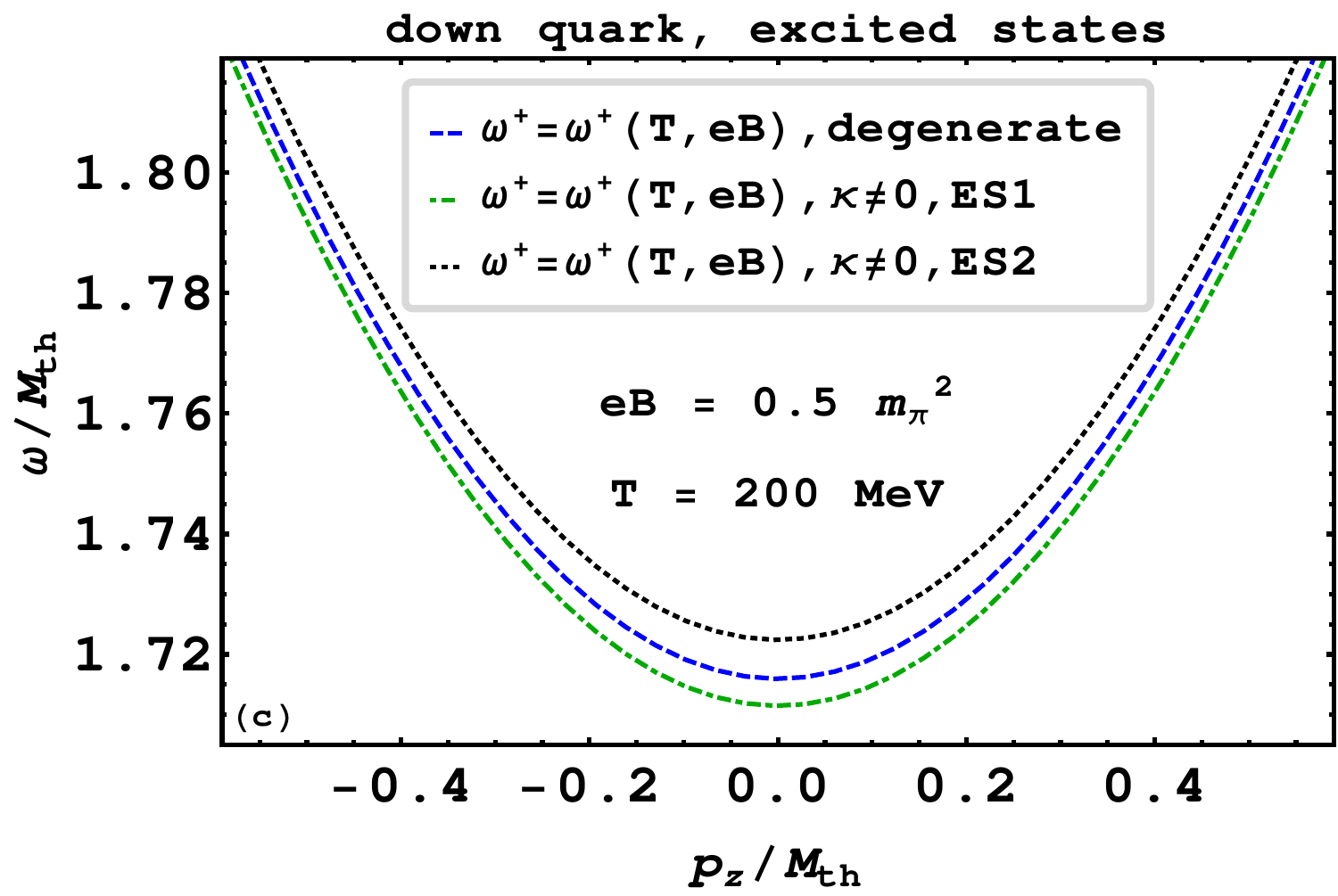} ~~ 
	\includegraphics[scale=0.34]{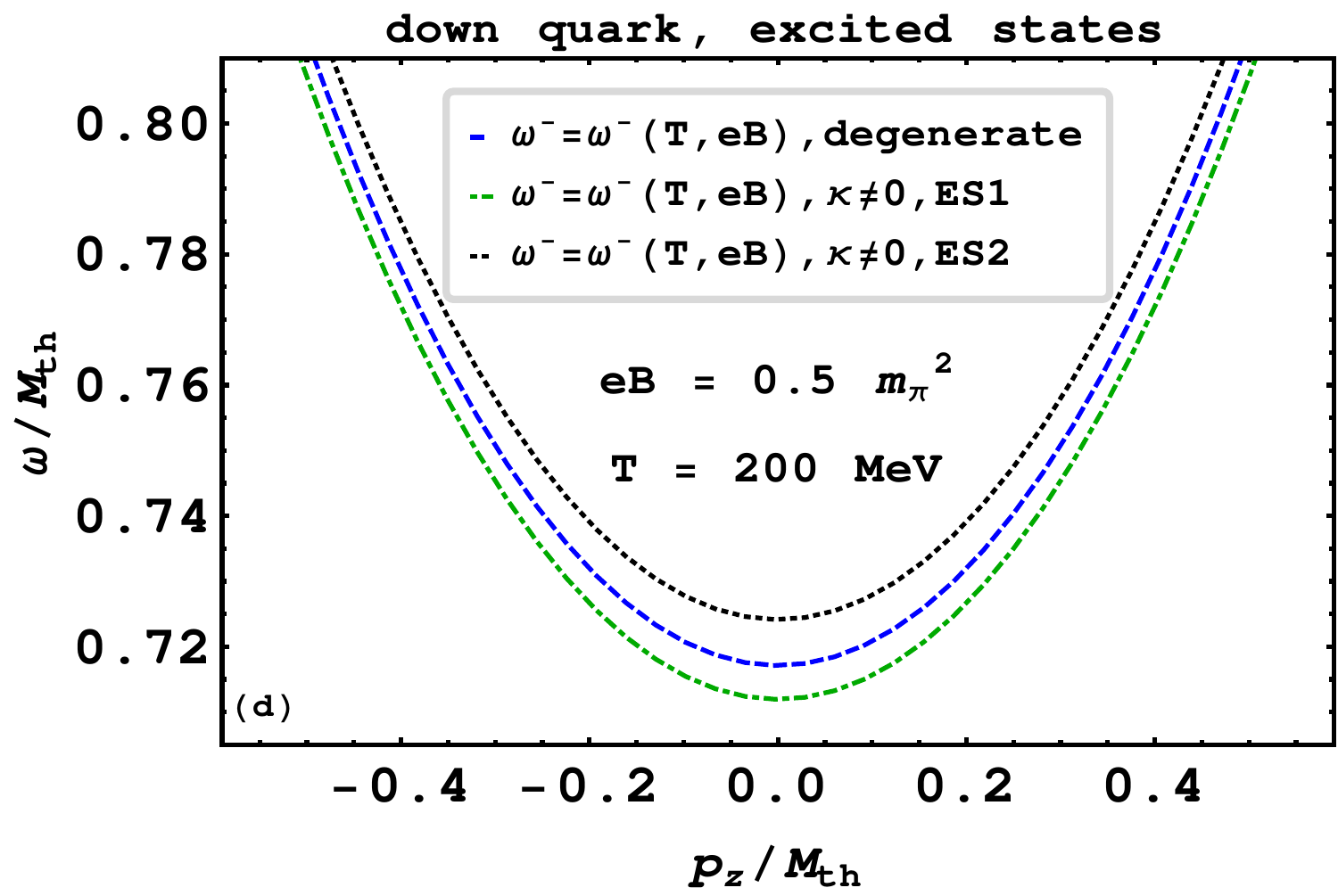}
	\caption{Dispersion of down quark in a hot magnetized medium with finite values of AMM at $ T = 200 $ MeV and $ eB = 0.5~m_\pi^2 $ is depicted in the weak field limit. In the left (right) panel, we have illustrated the particle (hole) like excitations. In the top (bottom) panel, we have plotted the ground (excited) state. `ES1' and `ES2' refer to non-degenerate excited states with $\SB{n,s=1,-1}$ and $\SB{n,s= 0,1} $ respectively. Here $ \SB{n, s} $ labels the quantized values of the energy eigenvalues of Dirac equation with finite values of AMM as given in Eq.~\eqref{Eq_Dispersion}.}
	\label{Fig_Ddown}
\end{figure}
\begin{figure}[hbt]
	\includegraphics[scale = 0.34]{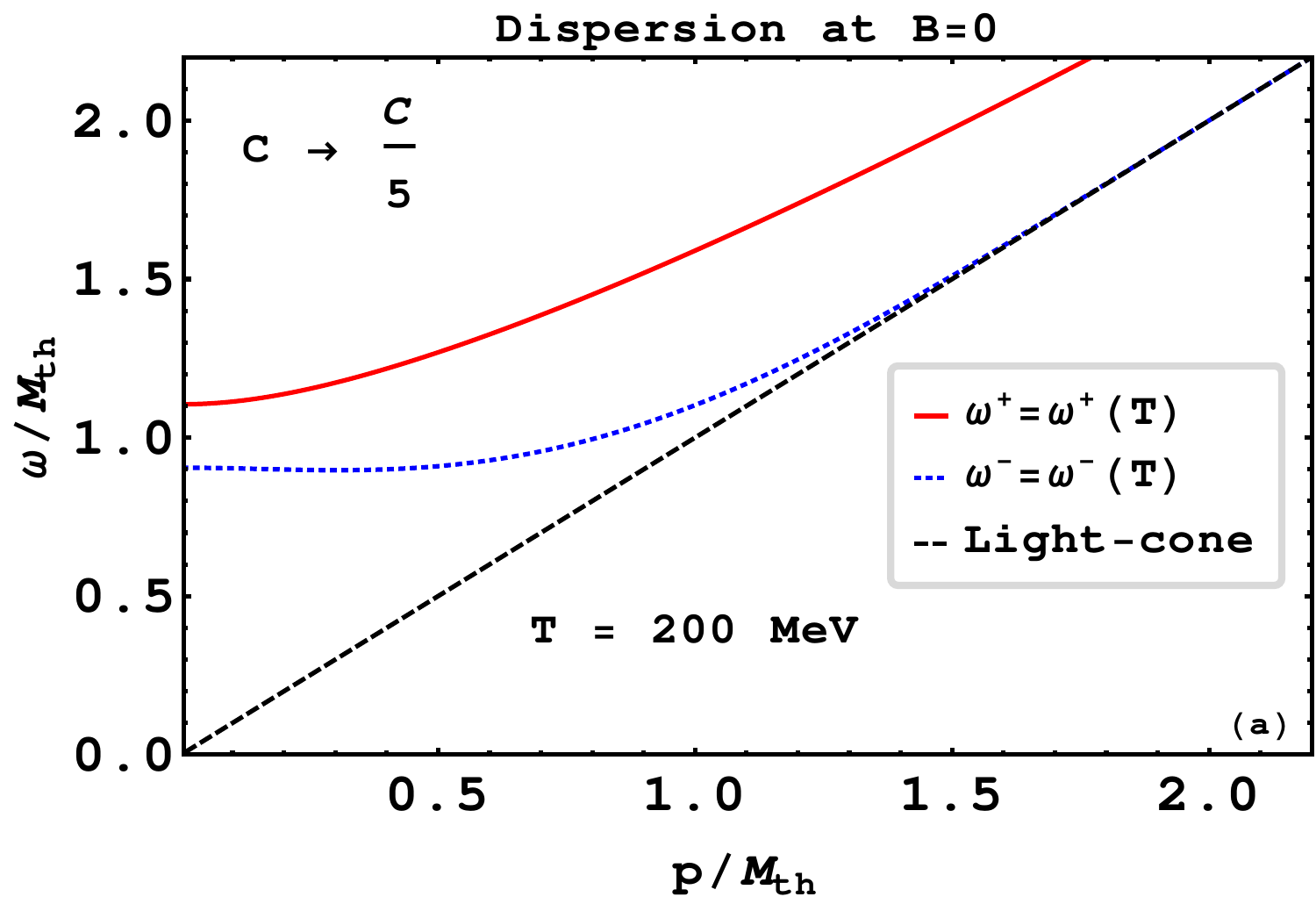}~~
	\includegraphics[scale = 0.34]{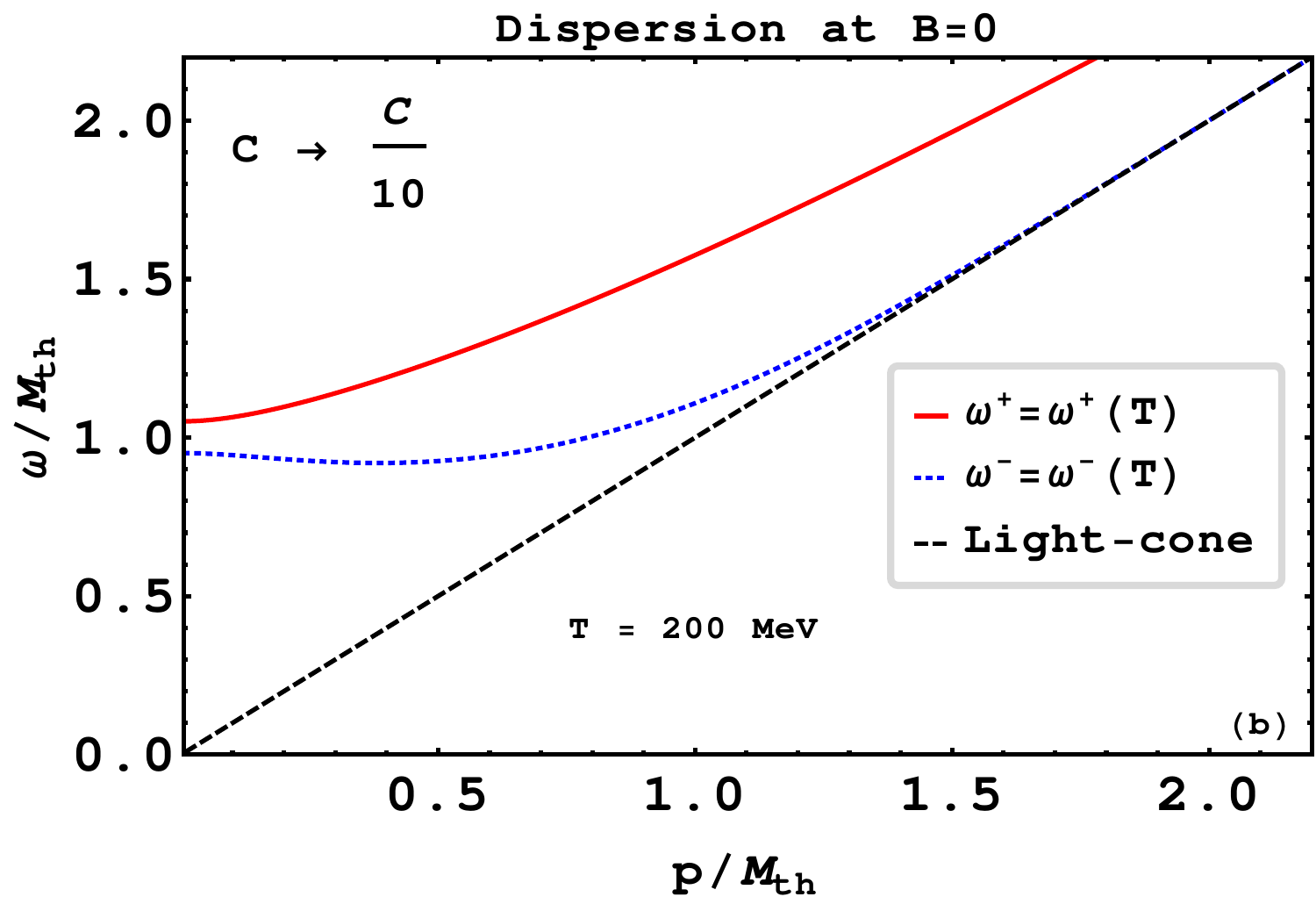} \\
	\includegraphics[scale = 0.34]{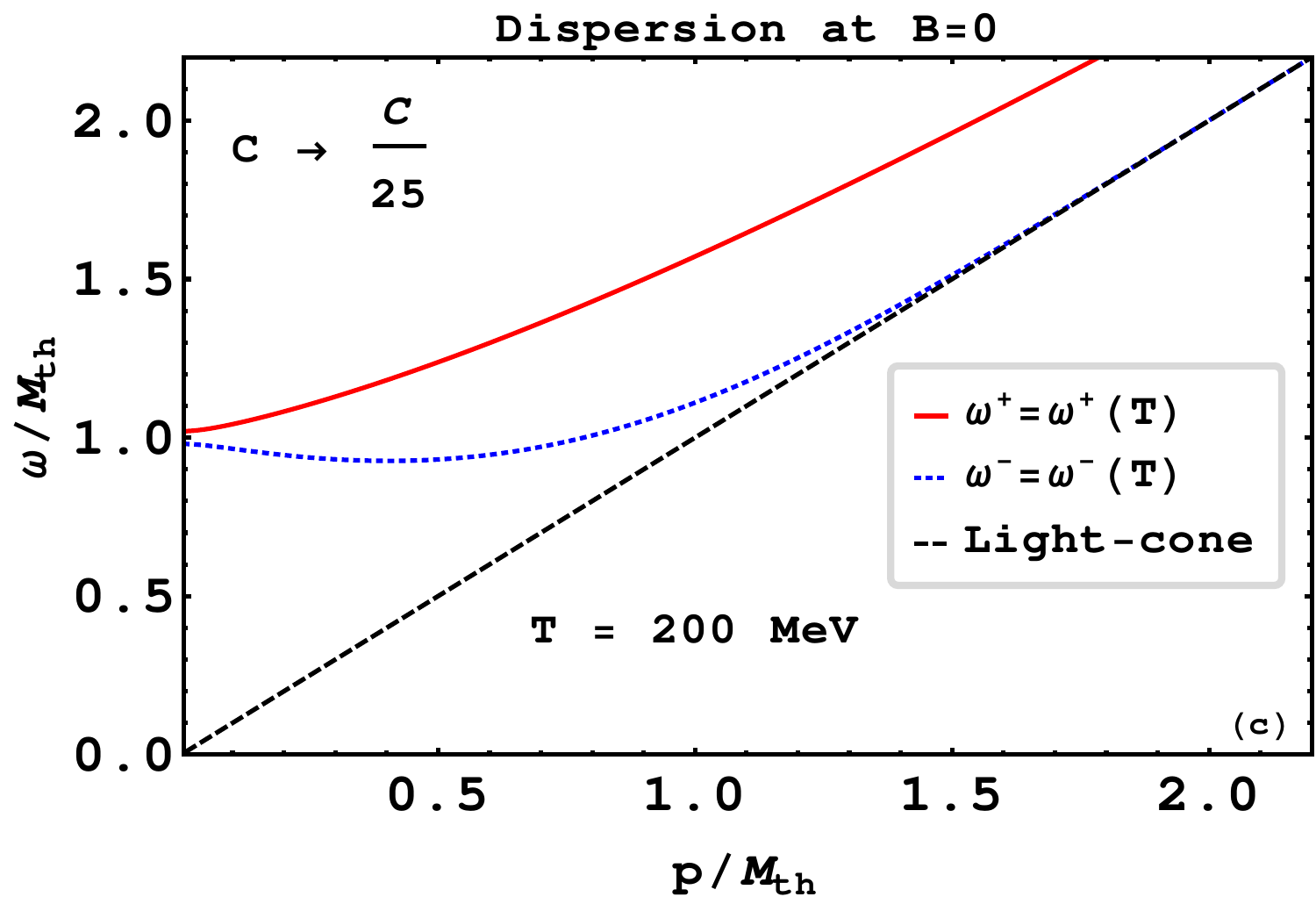}~~
	\includegraphics[scale = 0.34]{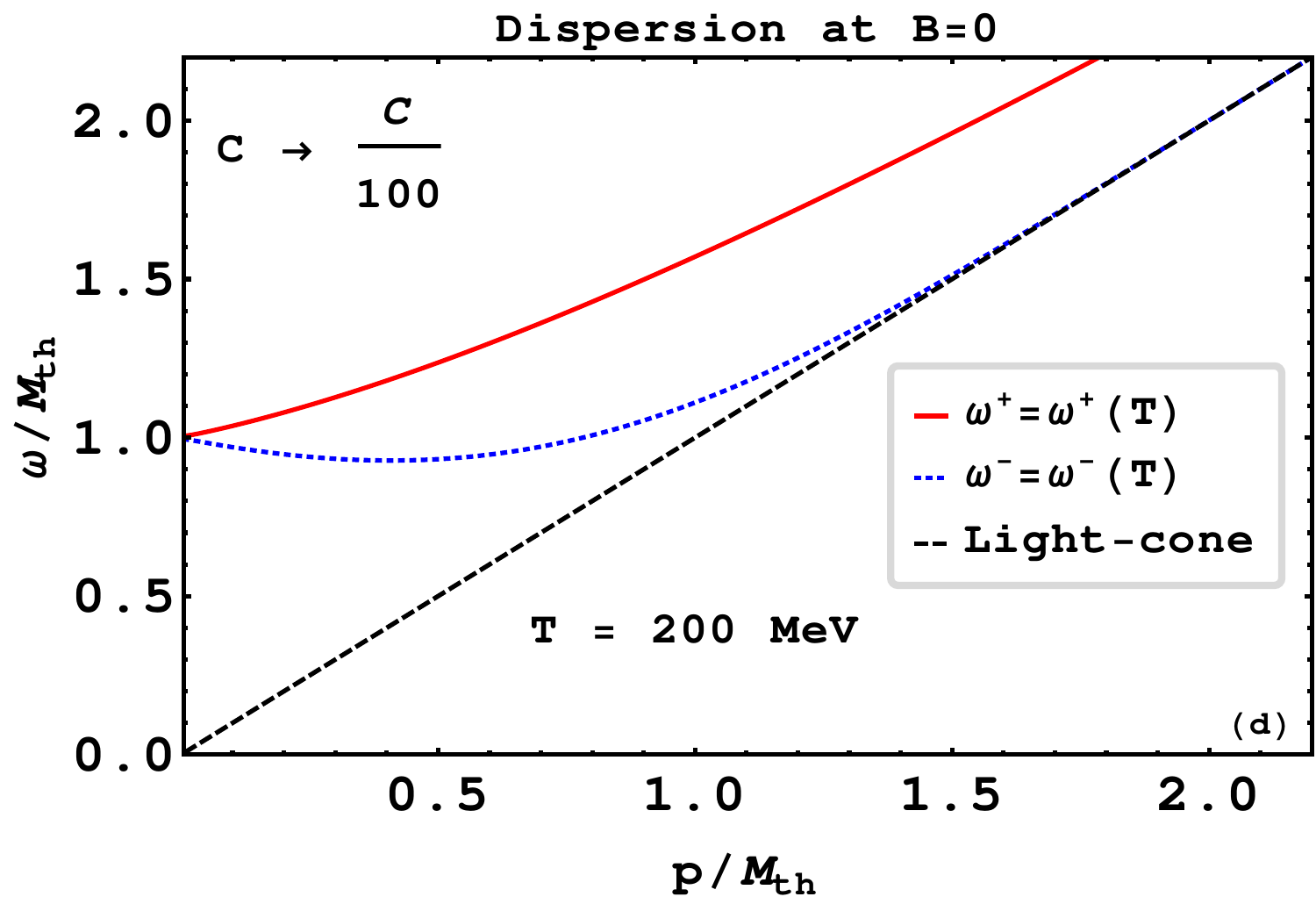}
	\caption{Collective modes of a fermion with explicit chiral symmetry breaking in an isotropic plasma for different values of $ C $ defined in Eq.~\eqref{HTL_C}. The solid red line represents the particle-like excitation, while the blue dotted line represents the hole-like excitation. For very small values of $ C $ we recover the well known results in the chiral limit.}
	\label{Fig_C0limit}
\end{figure}

Before showing the results in presence of background magnetic field, we first discuss dispersive properties of a massive  fermion   in a strongly interacting thermal medium. In Fig.~\ref{Fig_DeB0} we have shown the quasi-particle dispersion modes of a fermion at finite temperature when the chiral symmetry is explicitly broken. For comparison free particle dispersion relation with thermal mass $ \Mth  $ is also plotted. To get these results one has to solve the magnetic field independent versions of Eqs.~\eqref{HTL_Dp} and \eqref{HTL_Dm} for the poles which are given by~\cite{Weldon:1982bn}
 \begin{align}
 	D_+(p^0,p) &= \SB{(1+a) p^0 +b } - \sqrt{(1+a)^2 p^2 +C^2}=0 \label{HTL_DTp}\\
 	 \text{and}~~~~~~~~~	D_-(p^0,p) &= \SB{(1+a) p^0 +b } + \sqrt{(1+a)^2 p^2 +C^2}=0 \label{HTL_DTm}~.
 \end{align}
Note that in both the equations above, if $ \omega $ is a solution, so is $ -\omega $. Here only positive energy solutions are plotted. As evident from the plot, the presence of finite temperature significantly modifies the dispersive property of the fermion as its collective oscillation splits into two different modes.  The solution of Eq.~\eqref{HTL_DTp}, represented by red solid line and denoted as `$\omega^+ $', corresponds to the in-medium propagation of a particle-like excitation. This is because it will resemble the free particle dispersion if the medium effects are turned off (i.e. structure factors $ a,b \to 0 $). On the other hand, the mode shown by the blue dotted line and labeled as `$ \omega^- $', is purely a medium effect which we get by solving Eq.~\eqref{HTL_DTm}. This mode corresponds to hole-like excitation. These results are consistent with previous observations in Refs.~\cite{Petitgirard:1991mf,Haque:2018eph}. Furthermore, following Ref.~\cite{Petitgirard:1991mf}, one can derive the analytical results for the dispersion relations given by Eqs.~\eqref{HTL_DTp} and \eqref{HTL_DTm} for small values of $ p $. They are given by
\begin{align}
	\omega^+ &= M_+ + \frac{1}{\Mth^2+ M_+^2} \SB{ \frac{\Mth^2}{3 M_+} + \frac{1}{2m} \FB{ \frac{3 M_+^2 -\Mth^2}{3 M_+} }  } p^2 + \mathcal{O}(p^4) \label{Omega_p}
	\\
	\text{and}~~~~~	\omega^- &= M_- + \frac{1}{\Mth^2+ M_-^2} \SB{ \frac{\Mth^2}{3 M_-} + \frac{1}{2m} \FB{ \frac{3 M_-^2 -\Mth^2}{3 M_-} }  } p^2 + \mathcal{O}(p^4) \label{Omega_m}	
\end{align}
where $	M_\pm  = \frac{\sqrt{m^2 + 4\Mth^2} \pm m}{2}$. To obtain these results, one has to note the fact that the expression for $ \alpha $ given by Eq.~\eqref{HTL_alpha} does not contain any $ T $-dependent term and therefore can be neglected in the first approximation. Note that, under these assumptions, the gap at $ p = 0 $ between particle and hole-like excitations is exactly given by $ m $. As the mass parameter $ m $ increases from small to large values, the coefficient of the $ p^2 $ term in Eq.~\eqref{Omega_m} changes from negative to positive. Consequently for extremely light masses, this collective mode branch shows a minimum~\cite{Weldon:1982bn,Bellac:2011kqa,Laine:2016hma,Strickland:2019tnd,Mustafa:2022got}. One can easily check this numerically by taking the limit $ C\to0 $ which will be discussed later in the Sec.~\ref{B0Limit}. However, as the mass $ m $ increases, this minimum vanishes, and the solution transforms into a monotonic function of $ p $. 
Now, we study the collective modes in the limit $ \Mth \ll p \ll T $ and analytical expression for particle and hole-like excitations are given by~\cite{Petitgirard:1991mf}
\begin{align}
\omega^+ &= p +\dfrac{\Mth^2}{p} + \dfrac{m^2}{2p} +\cdots \label{Omega_p2}, \\
\omega^- &= p + 2p \exp\TB{- \FB{ \frac{2p^2}{\Mth^2} +1} } \exp\TB{ -\frac{m^2}{\Mth^4}p^2 } +\cdots ~\label{Omega_m2}.
\end{align}
In the chiral limit, these expressions are same as obtained in Refs.~\cite{Weldon:1989ys,Mustafa:2022got}. Notably, for large momenta the particle-like excitations closely resemble free particles in vacuum. Conversely, in the solution for hole-like excitations given by Eq.~\eqref{Omega_m2}, we observe the presence of the first exponential term which is akin to the massless case at large momenta. However, this is now accompanied by a new exponential factor that significantly accelerates the asymptotic behaviour as the mass parameter $ m $ increases.
This phenomenon becomes more apparent on comparing Fig.~\ref{Fig_DeB0} and Fig.~\ref{Fig_C0limit} (c), where in the former case the hole-like excitation approaches the light-cone for smaller  values of momentum compared to the latter case as discussed later.
 Finally, at small momenta both collective modes are equally important, but at large momenta the collective mode corresponding to the hole-like excitations decouples from the plasma.   
  
Now we move on to show the results for the scenario when finite background field is present. Since we are working in the weak field limit, we will assume $ eB = 0.5 $~$ m_\pi^2 $ throughout the rest of the discussions. Moreover, we will consider $ q_u = 2/3 e $ for the charge and $ \kappa_u = 0.29 $~GeV$ ^{-1} $ as a parametric value of the AMM of up quark. Similarly we will choose $ q_d = -1/3 e $ and $ \kappa_d = 0.36 $~GeV$ ^{-1} $ for down quarks. These values for AMM of up and down quarks are considered following constituent quark model such that they reproduce the AMMs of proton and neutron~\cite{Fayazbakhsh:2014mca}.

In Figs.~\ref{Fig_Dup} (a)$ - $(d) we have presented in-medium collective modes of an up quark in the presence of a background magnetic field considering the finite values of the AMM at $ T = 200 $~MeV. To arrive at these results we have solved Eqs.~\eqref{HTL_Dp} and \eqref{HTL_Dm} in a self-consistent manner. In all the plots particle (hole) like solutions are shown in the left (right) panel. It is well known that, in the presence of external magnetic field the energy eigenvalues of Dirac equation with finite values of AMM can be expressed as~\cite{Fayazbakhsh:2014mca,Chaudhuri:2021lui,Xu:2020yag}
\begin{equation}\label{Eq_Dispersion}
E_{nfs}^2 = p_z^2 + \fb{ \sqrt{m^2 + (2n+1 -s \xi_f )|q_f eB|} - s \xi_f k_f |q_f eB| }^2  
\end{equation}
where $ \xi_f = \sgn{q_f} $.  Note that from Eq.~\eqref{Eq_Dispersion} it is evident that the ground state of a positively (negatively) charged fermion is $ n=0 $ and $ s = 1 $ ($ s = -1 $) which is well known result for Landau quantization of a charged particle in a magnetic field.
Hence, it becomes evident that the introduction of a finite AMM has the effect of lifting the degeneracy that exists in the excited states in the absence of AMM. Specifically, it lifts the degeneracy that exists among the doubly degenerate excited states. Now to evaluate the dispersive properties of low lying quarks in a magnetized medium one needs to solve Eqs.~\eqref{HTL_Dp} and \eqref{HTL_Dm}. For this one requires the knowledge of the momentum component transverse to the magnetic field which is Landau quantized and can be expressed as
\begin{equation}\label{Eq_PT}
	-P_\perp^2 = p_x^2 +p_y^2 = \FB{ \sqrt{m^2 + (2n+1 -s \xi_f )|q_f eB|} - s \xi_f k_f |q_f eB| }^2 - m^2~.
\end{equation} 
Fig.~\ref{Fig_Dup} (a) shows the dispersive properties of the particle-like excitation i.e. solution of Eq.~\eqref{HTL_Dp} for up quark in a hot magnetized medium in the ground state i.e. for $ \SB{n,s = 0,1}$ with and without considering finite values of AMM. For comparison we have also plotted the in-medium ($ T\ne 0, eB = 0 $) dispersion for up quarks considering $ p=p_z $ in Eq.~\eqref{HTL_DTp} to explicitly show the effect of the presence of a background magnetic field. As evident from the figure, the results for thermo-magnetic medium are higher in magnitude compared to the thermal medium. Moreover, consideration of finite AMM of up quarks slightly decreases the particle-like solution in the whole range of momentum. Qualitatively similar effects are observed in case of hole-like excitation in the ground state as shown in Fig.~\ref{Fig_Dup} (b). Note that both the results for particle and hole-like excitations previously discussed are non-degenerate. We now turn our attention to the results for collective oscillations in the excited states. In Figs.~\ref{Fig_Dup} (c) and (d), we illustrate the in-medium dispersion relations for both particle-like and hole-like excitations of up quarks in the presence of a background magnetic field. 
As is apparent from the plots their magnitudes are clearly higher compared to the ground states.
In the absence of AMM, the results for particle and hole-like excitations, depicted by the blue dashed lines, exhibit a twofold degeneracy. These states correspond to $ \SB{n,s= 0,-1} $ and $ \SB{n,s=1,1} $. However, the introduction of AMM lifts this degeneracy. The green dot-dashed line corresponding to the quantum numbers $\SB{n,s=1,1} $, depicts the first excited state and has a slightly lower magnitude compared to the doubly degenerate excited states observed in the absence of AMM. Similarly, the black dotted line represents the second excited state with $\SB{n,s= 0,-1} $, and its magnitude is slightly higher than that of the degenerate excited states.

 In Figs.~\ref{Fig_Ddown} (a)$ - $(d) we have displayed the in-medium collective modes of a down quark in the presence of a background magnetic field taking into account finite values of the AMM at $ T = 200 $ MeV. In the case of the ground states of particle and hole-like excitations shown in the upper panel, the results exhibit  qualitatively similar behaviour as observed in the case of up quarks. However, it is important to note that in this scenario, the ground state corresponds to the quantum number $ \SB{n,s = 0,-1} $. Additionally, due to the smaller magnitude of the charge of a down quark, the relative increase in the magnitude of the collective oscillations  in the magnetized medium, compared to the thermal medium, is less pronounced when compared to the up quark. The results for the collective oscillations of the down quarks in the excited states also exhibit a similar qualitative behaviour as observed in the case of up quarks. Here also the presence of finite AMM lifts the twofold degeneracy which is present in the excited states in the absence of AMM. However, it should be noted that in the case of down quarks the states with quantum numbers $ \SB{n,s = 1,-1} $ and $ \SB{n,s = 0,1} $, represented by the green dot-dashed line and black dotted line, respectively, correspond to the first and second excited states when finite values of AMM are considered. These two states are degenerate in the absence of AMM, as depicted by the blue dashed line.

 \subsection{Recovering the results in the chiral limit}\label{B0Limit}
 In Figs.~\ref{Fig_C0limit} (a)-(d), we have shown the collective oscillations for both particle and hole-like modes of a fermion in a thermal medium. This was achieved by solving Eqs.~\eqref{HTL_DTp} and \eqref{HTL_DTm} while considering various values of $ C $, as defined in Eq.~\eqref{HTL_C}. Ultimately, these results recover the well-known outcomes in the chiral limit. 
 These plots also highlight the presence of a critical value of $ m $ beyond which there is no longer a minimum in the hole-like excitation. Instead, it becomes a monotonic function of momentum $ p $, as previously argued in Ref.~\cite{Petitgirard:1991mf}.
  
%~~~~~~~~~~~~~~~~~~~~~~~~~~~~~~~~~~~~~~~~~~~~~~~~~~~~~~~~~~~~~~~~~~~~~~~~~~~~~~~~~~~~~~~~~~~~~~~~~~~~~~~~~~~~~~~~~~~~~~  
\section{Summary \& Conclusion} \label{sec.sum}
In this work, the general structure of the fermion self-energy at one loop order is obtained in a magnetized medium considering the finite values of AMM in the scenario when chiral symmetry is explicitly broken.  It is found that the self-energy of a thermo-magnetically modified massive fermion  contains five non-trivial structure factors. These structure factors have been evaluated in a weak magnetic field within the HTL approximation using the real time approach of thermal field theory which has not been explored previously. It is well known that the collective modes of a fermion at finite temperature leads to particle and hole-like excitations. It is observed that, at small momenta, both collective modes are equally important, whereas at large momenta, the collective mode associated with hole-like excitations decouples from the plasma. The dispersion in non-degenerate ground state shows that the magnitude is greater than that of the zero magnetic field case. This effect is more pronounced in the case of up quarks, mainly due to the higher magnitude of their charge compared to the down quarks. We have also investigated the effects of inclusion of finite values of AMM of the quarks. Firstly we see that the presence of finite AMM modifies the magnetic mass. In such a  scenario the most significant effect is observed in the dispersion relation of the excited states. For up quarks, which are positively charged, the states corresponding to the quantum numbers $ \SB{n,s= 0,-1} $ and $ \SB{n,s=1,1} $ are doubly degenerate in the absence of AMM. However, when finite values of AMM are taken into consideration, this degeneracy is lifted. The state with quantum numbers $ \SB{n,s=1,1} $ corresponds to the first excited state and has a slightly lower magnitude in energy compared to the doubly degenerate excited states observed in the absence of AMM. Similarly, the state represented by $ \SB{n,s= 0,-1} $ becomes the second excited state and its magnitude is slightly higher than that of the degenerate excited states.   This effects are observed in both particle and hole-like excitations. Qualitatively similar effects are also observed in the dispersion of down quarks. In this case, the ground state corresponds to the quantum number $ \SB{n,s = 0,-1} $. Additionally, the states with quantum numbers $ \SB{n,s = 1,-1} $ and $ \SB{n,s = 0,1} $ correspond to the first and second excited states when finite values of AMM are considered. These two states are degenerate in the absence of AMM.

\section*{Acknowledgement}
NC would like to express gratitude to Dr. Aritra Das, Prof. Munshi Golam Mustafa, and Dr. Nazmul Hauque for valuable discussions regarding the HTL method and related subjects. N.C., P.R. and S.S. are funded by the Department of Atomic Energy (DAE), Government of India. S.G. is funded by the Department of Higher Education, Government of West Bengal, India.

 \appendix
 
 \section{Factorization of $D$} \label{app}
 Using the definitions of $ L^\mu $ and $ R^\mu $ from Eqs.~\eqref{Def_L} and \eqref{Def_R}, one can write  
 \begin{align}
 	L^2 &= \fb{\mhA p^0 + \mhBm}^2 -\SB{ \fb{ \mhA p_z - \cp  }^2 - \mhA^2 P^2_\perp},\label{Def_L2} \\
 	R^2 &= \fb{\mhA p^0 + \mhBp}^2 -\SB{ \fb{ \mhA p_z + \cp  }^2 - \mhA^2 P^2_\perp}, \label{Def_R2} \\
 	L\cdot R &= \fb{ \mhA p^0 + b }^2- \mhA^2 p^2 -{\bp}^2 + {\cp}^2\label{Def_LR},
 \end{align}
 where $	\mhA = 1 +a$ and $\mathds{B}^\pm = b \pm \bp$.
 Using the explicit expressions for $ L^2,~R^2 $ and $ L\cdot R $ from Eqs.~\eqref{Def_L2}-\eqref{Def_LR} we can rewrite $ D $ from Eq.~\eqref{Def_D} as the following manner
 \begin{align}
 	D  &= \mtC^2 - \mtD^2 -\mtB^2 -\mtA^2 P^2 - 2 \mtA \mtB p^0 \nn \\
 	&=  \mtC^2 - \mtD^2 + \mtA^2 p^2 - \FB{\mtA p^0 +\mtB}^2  \nn \\
 	&= D_+(p^0,p) ~D_-(p^0,p)~,
 \end{align}
 where
 \begin{align}
 	D_\pm(p^0,p;eB) &=  \FB{\mtA p^0 +\mtB} \mp \sqrt{\mtC^2 - \mtD^2 + \mtA^2 p^2 }~.
 \end{align}
 In the above equations 
 \begin{align}
 	\mtA &= - 2\mhA  c ~,\\
 	\mtB &= - 2 b c ~,\\
 	\mtD &= 2 \TB{ \mhA \bp p^0 +b \bp - \mhA\cp p_z }~,\\
 	\mtC &= \mhA^2 P^2 + b^2 +{\bp}^2 -{\cp}^2 + c^2 + 2 \mhA b p^0~.
 \end{align}

 \bibliographystyle{apsrev4-1}
\bibliography{Reference} % (In the file reference.bib you can add bibtex script for Inspire HEP for example)

\end{document}